\documentclass[aps,pra,manuscript]{revtex4}
\usepackage{amssymb}
\usepackage{graphicx}
\usepackage{epsfig}
\usepackage{color}
\usepackage{bm}

\begin{document}
\draft
\title{Multi-timescale microscopic theory for radiation degradation of electronic and optoelectronic devices}

\author{Danhong Huang$^{1}$, Fei Gao$^{2}$ and D. A. Cardimona$^{1}$}
\address{$^{1}$Air Force Research Laboratory, Space Vehicles
Directorate, Kirtland Air Force Base, New Mexico 87117, USA\\
$^{2}$Department of Nuclear Engineering and Radiological Sciences, University of Michigan,\\
500 S. State Street, Ann Arbor, Michigan 48109, USA}

\date{\today}

\begin{abstract}
A multi-timescale hybrid model is proposed to study microscopically the degraded performance of electronic devices, covering three individual stages of radiation effects studies, including ultrafast displacement cascade, intermediate defect stabilization and cluster formation, as well as slow defect reaction and migration. Realistic interatomic potentials are employed in molecular-dynamics calculations for the first two stages up to $100$\,ns as well as for the system composed of layers with thickness of hundreds times of lattice constant. These quasi-steady-state results for individual layers are input into a rate-diffusion theory as initial conditions to calculate the steady-state distribution of point defects in a mesoscopic-scale layered-structure system, including planar biased dislocation loops and spherical neutral voids, on a much longer time scale. Assisted by the density-functional theory for specifying electronic properties of point defects, the resulting spatial distributions of these defects and clusters are taken into account in studying the degradation of electronic and optoelectronic devices, e.g., carrier momentum-relaxation time, defect-mediated non-radiative recombination, defected-assisted tunneling of electrons and defect or charged-defect Raman scattering as well. Such theoretical studies are expected to be crucial in fully understanding the physical mechanism for identifying defect species, performance degradations in field-effect transistors, photodetectors, light-emitting diodes and solar cells, and in the development of effective mitigation methods during their microscopic structure design stages.
\end{abstract}
\pacs{PACS:}
\maketitle

\section{Introduction}
\label{sec1}

Point defects (vacancies and interstitial atoms) are produced by the displacements of atoms from their lattice sites,\,\cite{book-gary,book-sigmund} where the atom displacements are mainly induced by a primary knockout atom (PKA) on a time scale shorter than $50$\,ps. This initial phase is followed subsequently by a defect reaction (clustering or dissolution of clusters),\,\cite{gao6} and further by the thermally-activated migration\,\cite{gao5} of the point defects and defect clusters in a time scale longer than $100$\,ns. The combination of all these processes, resulting in a significant concentration of surviving defects in the crystal, is physically termed {\em particle irradiation displacement damage} (in addition to the well know $\gamma$-ray electron ionization cascade damage). Such radiation displacement damage effects depend not only on the energy-dependent flux of the incident particles (protons, neutrons, ions, etc.) but also on the differential energy transfer cross sections (probabilities) for collision between atoms, interatomic coulomb interactions and kinetic-energy loss to electrons inside an atom. Irradiation temperature also significantly affects the motion of defects, their stability as clusters and the formation of Frenkel pairs.\,\cite{gao7}
\medskip

On the other hand, electron devices are usually classified either as electronic ones, where electrons respond to an applied voltage as a current flow, or optoelectronic ones, in which electrons perform interband/intraband optical transitions in the presence of an incident light signal.\,\cite{book-huang} For an electronic device (e.g., field-effect transistors in an integrated circuit), the momentum-relaxation time of electrons, due to scattering by randomly-distributed defects, plays a crucial role in determining the electron mobility,\,\cite{marshall} while the photo-excited electron lifetime, due to non-radiative recombination with defects, is proven to be a key factor affecting the sensitivity or the performance of optoelectronic devices (e.g., photodetectors and light-emitting diodes).\,\cite{anderson}
\medskip

In a perfect crystal, the continuous free-electron states are quantized into many Bloch bands separated by energy gaps, and these Bloch electrons move freely inside the crystal with an effective mass different from that of the free electrons.\,\cite{book-callaway} In the presence of defects, however, the field-driven current flow of Bloch electrons in a perfect crystal will be scattered locally by these defects, leading to a reduced electron mobility. In addition, photo-excited Bloch electrons could acquire a shortened lifetime, giving rise to a degraded quantum efficiency due to enhanced non-radiative recombination with defects. The dangling bonds attached to the point defects may capture extra electrons to form charged defects. In this case, the positively-charged holes in the system will be trapped to produce a strong space-charge field, while the negatively-charged electrons may generate the so-called $1/f-$current noise in their chaotic motion due to the presence of many potential minima and maxima from randomly-distributed charged defects.
\medskip

\begin{figure}[p]
\centering
\epsfig{file=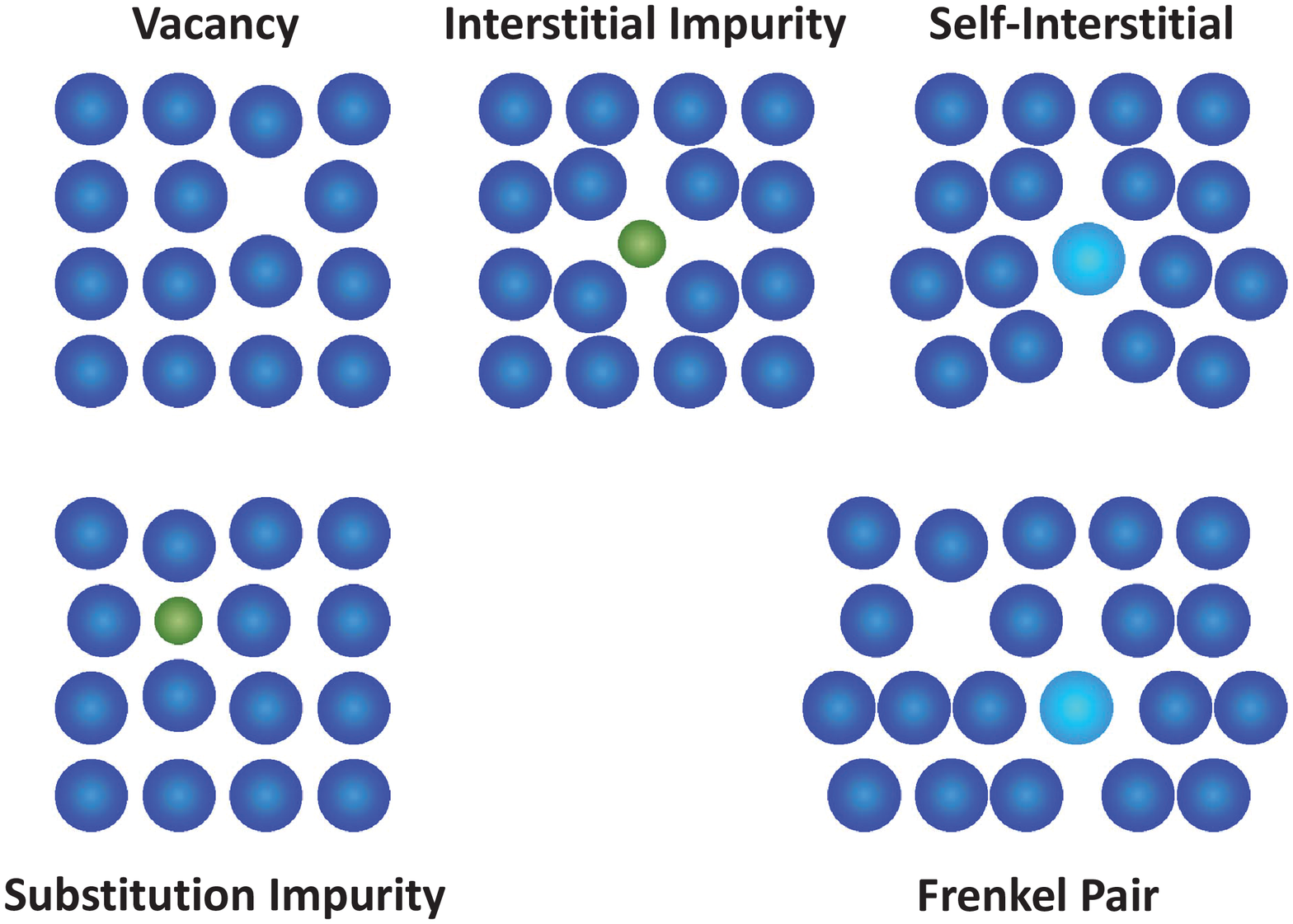,width=0.7\textwidth}
\caption{Two-dimensional illustrations of different types of point defects in a crystal.}
\label{f1}
\end{figure}

Point defects in crystals, as shown in Fig.\,\ref{f1}, can be generated by particle irradiation in both bulk and nano-crystals composed of many grains with different sizes.\,\cite{gao9} One of the effective calculation methods for studying the non-thermal spatial-temporal distributions of radiation-induced point defects is the molecular-dynamics (MD) model based on a stepped time-evolution approach (also termed the collisional and thermal spike stages), which involves the total force by summing over the interatomic potentials from all the atoms in a finite system.\,\cite{gao2} The lattice vibration at finite temperatures can be taken into account by an initial thermal-equilibrium state for atoms (intrinsic vacancies and interstitials) in the system plus an initial velocity for one of the atoms in a specific direction. The system size increases quadratically with the initial kinetic energy of particles and the time scale runs up to several hundred
picoseconds (called the quenching stage). Therefore, the defect reaction process by thermal migration cannot be included in this MD model due to its much longer time scale, although the other processes, such as displacement of lattice atoms, energy dissipation, spontaneous recombination and clustering, can be fully taken into account. If the system time evolution goes beyond $100$\,ps, the kinetic lattice Monte-Carlo method can also be used.\,\cite{gao1} However, if the time scale exceeds several hundreds of nanoseconds (also called the annealing stage), the rate theory\,\cite{ryazanov,stoller} has to be called in for studying the steady-state properties of the surviving defects (up to hours or days or even months).
\medskip

We know that the MD model with a realistic interatomic potential has been developed for studying the non-thermal spatial-temporal distributions of radiation-induced point defects in noble transition metals and alloys, and the density-functional theory has been widely used for calculating electronic properties of defects with pre-assumed specific defect configurations. On the other hand, a quantum-mechanical model has been well established for investigating defect effects on semiconductor electronic devices in the presence of spatially-uniform and randomly distributed point defects. However, to the best of our knowledge, no first-principle model and theory has been proposed so far to study microscopically the degraded performance of electronic devices induced by particle irradiation displacement damage. Therefore, the theory presented in this paper is expected to be very important in understanding the full mechanism for characterizing defects, performance degradations in transistors, photodetectors, light-emitting diodes and solar cells, as well as in developing effective mitigation in early design stages. Equipped with our current multi-timescale microscopic theory, at one end of the problem, the experimental characterization of post-irradiated space-based devices allows us to correctly extract useful information about particle irradiation sources. From the other end of the problem, this also lets us predict reliably the accumulated performance degradation of devices with time based on space-weather forecast after a satellite has passed through the radiation belts many times.
\medskip

Some of the equations presented below will be well-known to researchers in materials science field, however, researchers in device physics field may not be aware of them. With this paper, we hope to bridge the gap between researchers studying radiation-induced damage in materials and researchers studying radiation-induced performance degradation in devices. This should allow the formalism developed for the investigation of radiation-induced structural defects in nuclear reactor materials\,\cite{book-gary} to be extended to the investigation of device performance degradation effects
induced by particle radiation found in space-based systems.
\medskip

The rest of the paper is organized as follows. In Sec.\ \ref{sec2}, we present our atomic-scale MD model to cover both the ultrafast defect generation and intermediate defect stabilization stages, as well as the mesoscopic-scale rate theory for defect migration and interaction processes.
In Sec.\ \ref{sec3}, master equations for both planar dislocation-loop and spherical void growth are introduced for studying surface and bulk sink dynamics, respectively. In Sec.\ \ref{sec4}, master equations are presented for exploring the steady-state spatial distribution of defects in layered structure materials. In addition, a density-functional theory is introduced for specifying electronic properties of point defects, and four device physics models are employed for characterizing and understanding defect-assisted resonant tunneling, reduced carrier mobility, non-radiative recombination with defects and inelastic light scattering by charged defects. Finally, some concluding remarks are presented in Sec.\ \ref{sec5}.

\section{Model and Theory}
\label{sec2}

\subsection{Atomic-Scale Modeling for Ultrafast Defect Generation \textmd{(displacement cascade: $t<100$\,ps)}}
\label{sec2.1}

\begin{figure}[p]
\centering
\epsfig{file=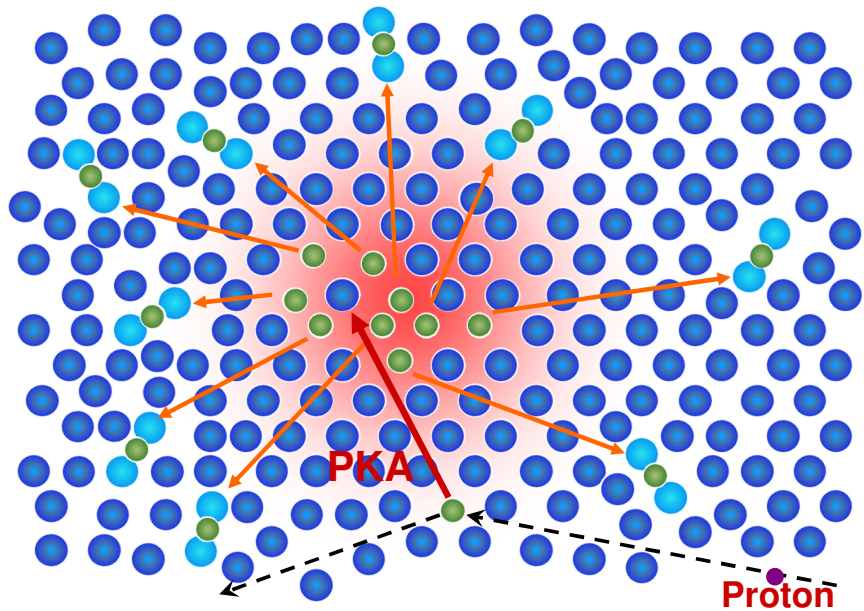,width=0.7\textwidth}
\caption{Two-dimensional schematic of a displacement cascade induced by incident protons on a crystal.}
\label{f2}
\end{figure}

A schematic of a displacement-cascade event by proton irradiation is shown in Fig.\,\ref{f2}. For the neutron-nucleus elastic collision, this process can be simply regarded as colliding hard spheres as an approximation due to their charge neutrality. The more complicated inelastic collision of neutrons with a nucleus, however, could involve generating an additional neutron [$(n,2n)$-process] or photon emission [$(n,\gamma)$-process], which are both important to the displacements of atoms. For the proton-nucleus elastic collision, on the other hand, the extra interaction (potential function) between the electron cloud and the proton should be considered.
\medskip

Commonly, the end product of the particle or neutron collision results in the PKA with an excess kinetic energy, and the subsequent atom-atom interaction represents the most fundamental physical mechanism of the radiation displacement damage.\,\cite{gao4,gao8} Since the radiation damage events are random in nature, a large number of damage events are required to obtain good statistics by choosing different directions and locations for PKA. On the other hand, the dynamics in the damage procedure can be accurately described by employing a realistic interatomic potential (MD model). For incident charged particles, the detailed form of the interatomic potential depends on the closest separation between two collision partners, which itself is determined by the kinetic energy of the incident particles (e.g., heavy-slow ions and relativistic electrons).
\medskip

The point defect generation as a result of displacement cascades is closely related to the PKA energy, which can be described statistically by an average transfer energy to the PKA. Such an average transfer energy can be calculated by using the energy-loss theory and measured by the so-called proton (electron) energy-loss spectroscopy as a function of various incident charged particle energies. The defects can also be identified experimentally by using positron annihilation.\,\cite{makkonen} With help from the computed energy loss of incident particles per unit length (called the loss function), the range of the particle before its full stop inside a crystal can be found. On the other hand, the MD method has been widely employed to simulate defect generation in a number of semiconductors, including Si,\,\cite{add1} SiC,\,\cite{gao4} GaAs,\,\cite{add2} and GaN.\,\cite{add3} These simulations provide important insights into the mechanisms for defect generation in semiconductors and predict the number and type of defects, spatial distribution of defects and initial correlation among defect species produced by the incident radiation for subsequent device level models.
\medskip

Basically, in MD simulations the time evolution of a set of interacting particles is tracked via the solution of Newton's equations of motion as shown below:

\begin{equation}
{\bf F}_j(t)=m_j\,\frac{d^2{\bf r}_j(t)}{dt^2}\ ,
\end{equation}
where the indices $j=1,\,2,\,\cdots,\,N$ label individual $N$ particles in the system, ${\bf r}_j(t)=[x_j(t),\,y_j(t),\,z_j(t)]$ is the position vector of the $j$th particle and $\displaystyle{{\bf F}_j(t)=-\sum_{k\ne j}\nabla_jV_{jk}}$ is the force acting upon the $j$th particle at time $t$ with interacting potential $V_{jk}$ between the $j$th and $k$th particle, and $m_j$ is the mass of the corresponding particle. In general, ${\bf F}_j(t)$ will depend on both particle positions and velocities at time $t$. To integrate the above second-order differential equations, the instantaneous forces acting on the particles and their initial positions and velocities need to be specified. Due to the many-body nature of the problem, the equations of motion have to be discretized and solved numerically. The MD trajectories are defined by both position vector ${\bf r}_j(t)$ and velocity vector $\displaystyle{{\bf v}_j(t)=\frac{d{\bf r}_j(t)}{dt}}$, and they describe the time evolution of the system in position-velocity phase space. Accordingly, the positions and velocities are propagated with a small time interval $\Delta t$ using numerical integrators. The numerical integration of Newton's equations of motion is to find an expression that defines positions ${\bf r}_j(t+\Delta t)$ at time $t+\Delta t$ in terms of the already known positions ${\bf r}_j(t)$ at time $t$. Because of its simplicity and stability, the Verlet algorithm is commonly used in MD simulations.\,\cite{add4} However, other popular algorithms, such as leapfrog, Velocity Verlet, Beeman's algorithms,\,\cite{add4,add5} predictor-corrector,\,\cite{add6} and symplectic integrators,\,\cite{add7} are also widely adopted. For non-PKA particles, their two initial conditions can be set as ${\bf r}_j(-\Delta t)={\bf r}_j(0)={\bf R}_j$, where ${\bf R}_j$ is the lattice vector for the $j$th site. If the PKA is given an initial velocity ${\bf v}_0$, in addition to ${\bf r}^{\rm PKA}(0)={\bf r}^{\rm PKA}_0$, this leads to another initial condition ${\bf r}^{\rm PKA}(-\Delta t)={\bf r}^{\rm PKA}_0-{\bf v}_0\Delta t$.
\medskip

In MD simulations, the atomic force field is crucial to determine physical systems in which collections of atoms are kept together by interatomic forces that can be calculated from empirical or semi-empirical interatomic potentials. Because of extensive applications of MD methods in materials science, a variety of techniques have been utilized over the years to develop reliable atomic-potential models. One of the early successful attempts to include many-body effects was the introduction of the embedding functional,\,\cite{add8} which depends nonlinearly upon the coordination number of each atom. This development leads to the birth of the embedded atom method (EAM),\,\cite{add9} which provides a relatively accurate description for noble transition metals as well as their alloys. However, the Tersoff potential formalism\,\cite{add10} is based on the concept of bond order and has been widely applied to a large number of semiconductors. Novel many-body forms have been tried in the attempt to capture as much as possible the physics and chemistry of the bonding. A typical analytical form is constituted by a number of functions, depending on geometrical quantities, such as distances or orientations, or on intermediate variables, such as atom coordinations. For example, a Tersoff potential has the appearance of a pair potential as below:

\begin{equation}
V=\frac{1}{2}\,\sum_{i,j=1}^N\,V_{ij}=\frac{1}{2}\,\sum_{i,j=1}^N\,\phi_{\rm R}(r_{ij})+\frac{1}{2}\,\sum_{i,j=1}^N\,B_{ij}\,\phi_{\rm A}(r_{ij})\ ,
\label{newton4}
\end{equation}
where the terms with $i=j$ are excluded in the above summations, $r_{ij}=|{\bf r}_i-{\bf r}_j|$ and the first and second terms represent repulsive and attractive interactions, respectively. However, the second term in Eq.\,(\ref{newton4}) is not a true pair potential since $B_{ij}$ is not a constant. In fact, it is the bond order for the bond joining the $i$th and $j$th atoms, and it is a decreasing function of a ``coordination'' $G_{ij}$ assigned to the bond. Therefore, we have $B_{ij}=B(G_{ij})$ and $G_{ij}$ is in turn defined by

\begin{equation}
G_{ij}=\sum_k\,f_{\rm c}(r_{ik})\,g(\theta_{ijk})\,f_{\rm c}(r_{jk})\ ,
\end{equation}
where $f_{\rm c}(r)$ and $g(\theta)$ are suitable functions. The basic idea is that the $i$-$j$ bond is weakened by the presence of other $i$-$k$ and $j$-$k$ bonds involving the intermediate atom at site $k$. The amount of bond weakening is determined by where the other bonds are. Angular terms appear necessary to construct a realistic model. When using a potential, the simulator should always be familiar with its transferability properties, and validate critically the results obtained in unusual conditions, for example, for very low coordinations, very high temperature, or very high pressure.

\subsection{Atomic-Scale Modeling for Intermediate Defect Stabilization \textmd{(stable defect and cluster formations: $100$\,ps$<t<10$\,ns)}}
\label{sec2.2}

The atom-displacement-generated point defects (vacancies and interstitial atoms) under particle irradiation will thermally diffuse in space, and interact and react with dynamical distributed bulk sinks, planar dislocation loops,\,\cite{book-lothe} spherical voids and clusters (due to collision cascade) at the same time.\,\cite{book-gary} Generally, the kinetic energy of the incident particles (or equivalently, the recoil energy of the struck atom) determines the specie and number of individual point defects during the initial phase (in addition to the rate of defect generation), while the flux of the the incident particles decides the defect density and the nature of point-defect diffusion,\,\cite{book-mehrer} i.e., either in an independent way (for low-density non-interacting point defects) or in a direction-correlated way (for high-density interacting point defects).\,\cite{gao3}
\medskip

The macroscopic property changes of the irradiated system are related to the particle energy-flux per unit time by the so-called damage function which is extracted by experimental measurements. However, the damage function is found to depend on the initial approximation in a sensitive way. Therefore,
we are not able to treat physically the radiation displacement damage effects as a black box through a fitting procedure. Instead, we should understand the full dynamics of these defects on all the time scales after they have been produced. The spatial distribution of the mobile Frenkel pairs (i.e., vacancy-interstitial pairs) that are created is crucial in determining the number that survive annihilation or immobilization by clustering due to damage cascade.
\medskip

The statistically-averaged spatial distribution of point defects that are generated can be calculated based on the defect formation and recombination rates, as well as the follow-up processes for defect diffusion, interactions and reactions.\,\cite{gao5} If the degree of atom displacements is limited due to high incident particle kinetic energies and low number intensities, we generally seek the {\em radiation degradation effects} on electronic and optoelectronic devices rather than looking at radiation damage effects on the material level when there is a significant level of atom displacements under intense low-energy particle irradiation.\,\cite{marshall} This radiation degradation depends not only on the particle radiation source and material, but also on the device structure and functionality. The analytical theory below can only provide a qualitative understanding of the collision and thermally-activated diffusion processes, while the MD calculation based on a realistic interatomic potential is able to provide a quantitative conclusion for comparison with experimental data.

\subsubsection{Point-Defect Generation Rate}
\label{sec2.2.1}

The spatially-temporally-dependent damage rate per unit volume for the displacement atoms in a crystal can be calculated from\,\cite{book-gary}

\begin{equation}
{\cal G}_0({\bf r},\,t)=n_{\rm at}\int\limits_{E_{\rm min}}^{E_{\rm max}} d\varepsilon_{\rm i}\, \sigma_{\rm D}({\bf r}\vert \varepsilon_{\rm i})\,{\cal I}_{\rm ext}(t\vert \varepsilon_{\rm i})\ ,
\label{eqn-1}
\end{equation}
where $n_{\rm at}$ is the crystal atom volume density, ${\cal I}_{\rm ext}(t\vert \varepsilon_{\rm i})$ represents the external dynamical energy-dependent particle intensity per unit energy, $\sigma_{\rm D}({\bf r}\vert \varepsilon_{\rm i})$ stands for both the position- and energy-dependent {\em displacement} cross section, and $E_{\rm min}$ ($E_{\rm max}$) corresponds to the minimum (maximum) kinetic energy in the energy distribution of incident particles.
\medskip

Since the displacement cross section $\sigma_{\rm D}({\bf r}\vert \varepsilon_{\rm i})$ in Eq.\,(\ref{eqn-1}) physically describes the probability for the displacement of struck lattice atoms by incident particles, we can directly write down

\begin{equation}
\sigma_{\rm D}({\bf r}\vert \varepsilon_{\rm i})=\int\limits_{\varepsilon_1}^{\varepsilon_2} d\varepsilon_{\rm R}\, \sigma_{\rm C}({\bf r}\vert \varepsilon_{\rm i},\varepsilon_{\rm R})\,
{\cal N}_{\rm D}({\bf r}\vert \varepsilon_{\rm R})\ ,
\label{eqn-2}
\end{equation}
where $\sigma_{\rm C}({\bf r}\vert \varepsilon_{\rm i},\varepsilon_{\rm R})$ is the differential {\em energy transfer} cross section by collision, which measures the probability that an incident particle with kinetic energy $\varepsilon_{\rm i}$ will transfer a recoil energy $\varepsilon_{\rm R}$ to a struck lattice atom, ${\cal N}_{\rm D}({\bf r}\vert\varepsilon_{\rm R})$ represents the average number of displaced atoms due to collision, and $\varepsilon_1$ ($\varepsilon_2$) labels the minimum (maximum) recoil energy acquired by the struck lattice atom.
\medskip

Although ${\cal N}_{\rm D}({\bf r}\vert\varepsilon_{\rm R})$ can be directly determined by MD simulation, the simplest approximation to estimate the displaced atoms is the Kinchin-Pease model (for a solid composed of randomly arranged atoms by ignoring focusing and channeling effects). We can simply express ${\cal N}_{\rm D}({\bf r}\vert \varepsilon_{\rm R})$ as

\begin{equation}
{\cal N}_{\rm D}({\bf r}\vert \varepsilon_{\rm R})=\left\{
\begin{array}{llll}
0\ , & \mbox{for $\varepsilon_{\rm R}<E_{\rm th}({\bf r})$}\\
1\ , & \mbox{for $E_{\rm th}({\bf r})<\varepsilon_{\rm R}<2E_{\rm th}({\bf r})$}\\
\displaystyle{\frac{\varepsilon_{\rm R}}{2E_{\rm th}({\bf r})}}\ , & \mbox{for $2E_{\rm th}({\bf r})<\varepsilon_{\rm R}<E_{\rm c}({\bf r})$}\\
\displaystyle{\frac{E_{\rm c}({\bf r})}{2E_{\rm th}({\bf r})}}\ , & \mbox{for $\varepsilon_{\rm R}\geq E_{\rm c}({\bf r})$}
\end{array}\right.\ ,\
\label{eqn-3}
\end{equation}
where $E_{\rm th}({\bf r})$ is the displacement threshold energy which depends on the chemical-bond strength of the specific struck atom, and $E_{\rm c}({\bf r})$ represents the cut-off energy due to energy loss by electron stopping (i.e., excitation or ionization of internal electrons) which is generally related to the electronic states of the individual struck atom. In principle, the displacement threshold energy $E_{\rm th}({\bf r})$ can be calculated by using the atomic-scale theory for given interatomic potential, crystal direction and crystal structure.
\medskip

In addition, the magnitude of the differential energy transfer cross section $\sigma_{\rm C}({\bf r}\vert \varepsilon_{\rm i},\varepsilon_{\rm R})$ introduced in Eq.\,(\ref{eqn-2}), which can be regarded as the crystal response to the external particle collision with lattice atoms, depends on the detailed collision mechanism and the form of the scattering potential as well. Here, as a simple example, we first give the expression of the differential energy transfer cross section for the elastic scattering. For the well-known Rutherford elastic scattering model based on an unscreened Coulomb potential $U_{\rm R}(\rho)=Z_1Z_2e/\epsilon_0\rho$ for protons with $\rho$ being the radius in the local frame centered on the lattice atom, we get

\begin{equation}
\sigma_{\rm C}({\bf r}\vert \varepsilon_{\rm i},\varepsilon_{\rm R})\equiv\sigma_{\rm R}({\bf r}\vert\varepsilon_{\rm i},\varepsilon_{\rm R})
=\frac{\pi b_0^2({\bf r})}{4}\,\frac{\varepsilon_{\rm i}\gamma({\bf r})}{\varepsilon^2_{\rm R}}\ ,
\label{eqn-4}
\end{equation}
where $\gamma({\bf r})=4mM({\bf r})/[M({\bf r})+m]^2$, $m$ [$M({\bf r})$]
is the mass of the incident particle (surface atoms or different lattice atoms), $b_0({\bf r})=Z_1Z_2({\bf r})\,e^2/\eta({\bf r})\epsilon_0\varepsilon_{\rm i}$ with $Z_1$ and $Z_2({\bf r})$ being the nuclear charge numbers for particles and different lattice atoms, and $\eta({\bf r})=m/[M({\bf r})+m]$.
\medskip

If the kinetic energy of incident particles is very high, the Rutherford scattering model becomes no longer applicable. In this case, we have to consider hard-sphere type collision for neutrons, which leads to

\begin{equation}
\sigma_{\rm C}({\bf r}\vert \varepsilon_{\rm i},\varepsilon_{\rm R})\equiv\sigma_{\rm HS}({\bf r}\vert\varepsilon_{\rm i},\varepsilon_{\rm R})
=\frac{\pi B^2}{\gamma({\bf r})\varepsilon_{\rm i}}\,\ln\left[\frac{A}{\eta({\bf r})\varepsilon_{\rm i}}\right]\ ,
\label{eqn-5}
\end{equation}
where a Born-Mayer potential $U_{\rm B-M}(\rho)=A\,\exp(-\rho/B)$ is employed.
\medskip

On the other hand, for the nucleus scattering with heavy-slow ions represented by a power-law interacting potential $U_{\rm I}(\rho)=(e/\epsilon_0a_0)\,(Z_1/Z_2)^{5/6}(a_0/\rho)^2$, this leads to

\begin{equation}
\sigma_{\rm C}({\bf r}\vert \varepsilon_{\rm i},\varepsilon_{\rm R})\equiv\sigma_{\rm I}({\bf r}\vert\varepsilon_{\rm i},\varepsilon_{\rm R})
=\frac{4E_a({\bf r})\,a^2({\bf r})\,\xi({\bf r})}{\gamma({\bf r})\,\varepsilon_{\rm i}^2[1-4\xi^2({\bf r})]^2\sqrt{X({\bf r})[1-X({\bf r})]}}\ ,
\label{eqn-6}
\end{equation}
where $X({\bf r})=\varepsilon_{\rm R}/\gamma({\bf r})\varepsilon_{\rm i}$, $\xi({\bf r})=\cos^{-1}[\sqrt{X({\bf r})}]/\pi$, $a({\bf r})=0.8853a_0/[Z_1Z_2({\bf r})]^{1/6}$
is the screening length with $a_0$ being the Bohr radius and $E_a({\bf r})=(e^2/\epsilon_0a_0)\,[Z_1/Z_2({\bf r})]^{7/6}/\eta({\bf r})$.
\medskip

Especially, for the incidence of relativistic light electrons, we have

\[
\sigma_{\rm C}({\bf r}\vert \varepsilon_{\rm i},\varepsilon_{\rm R})\equiv\sigma_{\rm e}({\bf r}\vert\varepsilon_{\rm i},\varepsilon_{\rm R})
=\frac{\pi Z_2^2e^4}{\epsilon_0^2m_0^2c^4}\,\frac{1-\beta_0^2}{\beta_0^4}
\]
\begin{equation}
\times\left\{1-\beta_0^2\frac{\varepsilon_{\rm R}}{E_2({\bf r})}+\pi\frac{\alpha_0({\bf r})}{\beta_0}\left[\sqrt{\frac{\varepsilon_{\rm R}}{E_2({\bf r})}}-\frac{\varepsilon_{\rm R}}{E_2({\bf r})}\right]\right\}
\frac{E_2({\bf r})}{\varepsilon_{\rm R}^2}\ ,
\label{eqn-7}
\end{equation}
where $\beta_0=v/c$ with $v$ being the velocity of incident electrons, $E_2({\bf r})=[2\varepsilon_{\rm i}/M({\bf r})c^2]\,(\varepsilon_{\rm i}+2m_0c^2)$, $m_0$ is the free-electron mass and $\alpha_0({\bf r})=Z_2({\bf r})/137$. Moreover, we have the relation $\beta^2_0=1-(m_0c^2/\varepsilon_{\rm i})^2\leq 1$ for the relativistic-particle velocity and kinetic energy.
\medskip

For isotropic inelastic scattering with an energy loss $Q_0$, on the other hand, we have the differential energy transfer cross section

\begin{equation}
\sigma_{\rm C}({\bf r}\vert \varepsilon_{\rm i},\varepsilon_{\rm R})\equiv\sigma^\prime_{\rm in}({\bf r}\vert \varepsilon_{\rm i},\varepsilon_{\rm R})=
\frac{\sigma_{\rm is}({\bf r}\vert \varepsilon_{\rm i},Q_0)}{\gamma({\bf r})\varepsilon_{\rm i}}\,\left[1+\frac{Q_0({\bf r})}{\varepsilon_{\rm i}}\,\frac{A({\bf r})+1}{A({\bf r})}\right]^{-1/2}\ ,
\label{eqn-8}
\end{equation}
where $A({\bf r})=M({\bf r})/m$, and $\sigma_{\rm is}({\bf r}\vert \varepsilon_{\rm i},Q_0)$ represents the isotropic differential energy transfer cross section for the resolved resonance in the center-of-mass frame.
\medskip

In addition, the incident-particle kinetic energy is no longer a constant if the particles are charged, e.g., protons and ions. In this case, we have $\varepsilon_{\rm i}\rightarrow\varepsilon_{\rm i}(z)=\left[\sqrt{\varepsilon_0}-\kappa z/2\right]^2$ with $\varepsilon_0$ being the incident-particle energy at the left boundary $z=0$, where a layered structure in the $z$-direction is assumed. As a result, we find that electronic stopping will dominate at short distance, while the elastic collisions will dominate near the end of the range.
\medskip

As an example, by using $\sigma_{\rm C}({\bf r}\vert \varepsilon_{\rm i},\varepsilon_{\rm R})$ from Eq.\,(\ref{eqn-4}) and ${\cal N}_{\rm D}({\bf r}\vert \varepsilon_{\rm R})$ from Eq.\,(\ref{eqn-3}), the displacement cross section from Eq.\,(\ref{eqn-2}) becomes

\begin{equation}
\sigma_{\rm D}({\bf r}\vert \varepsilon_{\rm i})\approx\left[\frac{\gamma({\bf r})}{4E_{\rm th}({\bf r})}\right]\sigma_s({\bf r}\vert \varepsilon_{\rm i})\ ,
\label{add-1}
\end{equation}
where $\sigma_s({\bf r}\vert \varepsilon_{\rm i})=[\pi b^2_0({\bf r})/4]\,[\gamma({\bf r})\varepsilon_{\rm i}/E_{\rm th}({\bf r})]$. Furthermore, by using the result in Eq.\,(\ref{add-1}), the displacement damage rate per unit volume from Eq.\,(\ref{eqn-1}) is

\begin{equation}
{\cal G}_0({\bf r},\,t)=n_{\rm at}\,\bar{\sigma}_s({\bf r})\left[\frac{\bar{\varepsilon}_{\rm i}({\bf r})\gamma({\bf r})}{4E_{\rm th}({\bf r})}\right]{\cal F}_{\rm ext}({\bf r},\,t)\ ,
\label{add-2}
\end{equation}
where $\bar{\sigma}_s({\bf r})$ and $\bar{\varepsilon}_{\rm i}({\bf r})$ are the average values with respect to the incident particle intensity per unit energy ${\cal I}_{\rm ext}(t\vert \varepsilon_{\rm i})$ in the energy range of $E_{\rm th}({\bf r})/\gamma\leq \varepsilon_{\rm i}\leq\infty$, ${\cal F}_{\rm ext}({\bf r},\,t)$ is the integrated external particle intensity for the same energy range and the term in the bracket is the number of Frenkel pairs produced per incident particle.

\subsubsection{Point-Defect Diffusion Coefficient}
\label{sec2.2.2}

Even in the absence of particle irradiation, there still exist some thermally-activated vacancies at room temperature in a crystal. In this case, the Helmholtz free-energy function in thermodynamics can be applied by assuming the volume of the crystal is a constant. In the presence of crystal defects, both the entropy $S$ and the enthalpy $H_p$ of a perfect crystal will be changed. A straightforward calculation gives the thermal-equilibrium numbers of vacancies $C^{\rm eq}_{\rm v}$ and interstitials $C^{\rm eq}_{\rm i}$ as follows:

\begin{equation}
C^{\rm eq}_{\rm v,i}=\exp\left(\frac{S_{\rm v,i}}{k_{\rm B}}\right)\exp\left(-\frac{E_{\rm v,i}}{k_{\rm B}T}\right)\ ,
\label{eqn-9}
\end{equation}
where $T$ is the temperature, $E_{\rm v}$ is the vacancy formation energy, which is smaller than the interstitial formation energy $E_{\rm i}$, and $S_{\rm v}$ ($S_{\rm i}$) is the change in entropy due to vibrational vacancy (interstitial) disorder.
\medskip

Diffusion of defects is driven by forces other than the concentration gradient of defects, such as stress or strain, electric fields, temperature, etc. The second Fick's law\,\cite{book-gary} directly gives rise to the following diffusion equation on the macroscopic scale

\begin{equation}
\frac{\partial C_{\rm v,i}({\bf r},\,t)}{\partial t}=-\nabla\cdot\left[D_{\rm v,i}({\bf r},t)\nabla C_{\rm v,i}({\bf r},\,t)\right]\ ,
\label{eqn-10}
\end{equation}
where $D_{\rm v}({\bf r},\,t)=D({\bf r},t\vert C_{\rm v})$ and $D_{\rm i}({\bf r},\,t)=D({\bf r},t\vert C_{\rm i})$ are called the diffusion coefficients for vacancy and interstitial atoms, respectively, and $c_{\rm v,i}({\bf r},\,t)=C_{\rm v,i}({\bf r},\,t)/{\cal V}$ is the defect concentrations with ${\cal V}$ being the volume of the system considered.
\medskip

In addition, by assuming a microscopic random walk for the diffusion process, we get the Einstein formula

\begin{equation}
D({\bf r})=D_0({\bf r})\,\exp\left[-\frac{E_{\rm ac}({\bf r})}{k_{\rm B}T}\right]=\frac{1}{6}\,\lambda_{\rm d}^2({\bf r})\,\Gamma({\bf r})\ ,
\label{eqn-11}
\end{equation}
where the temperature-independent part, $D_0({\bf r})$, is proportional to the Debye frequency ($\sim 10$\,THz) and is independent of defect concentration,
$E_{\rm ac}({\bf r})$ is the activation energy for thermal diffusion, $\lambda_{\rm d}({\bf r})$ is the diffusion length and $\Gamma({\bf r})$ is the defect jump rate.
\medskip

For tracer-atom diffusion, the random-walk model can not be used. Instead, the diffusion process becomes correlated, described by the Haven coefficient\,\cite{book-gary} $f({\bf r})$, and we get $D({\bf r})=f({\bf r})\,\lambda_{\rm d}^2({\bf r})\,\Gamma({\bf r})/6$, where $f({\bf r})<1$ depends on the crystal structure and the diffusion mechanism.
\medskip

The lattice-atom correlated diffusion coefficients $D_{\rm a}^{\rm v,i}({\bf r},\,t)$ by means of vacancy and interstitial are given by

\begin{equation}
D_{\rm a}^{\rm v,i}({\bf r},\,t)=f_{\rm v,i}({\bf r})\,D_{\rm v,i}({\bf r})\,C_{\rm v,i}({\bf r},\,t)\ ,
\label{eqn-12}
\end{equation}
which depend on the defect concentrations in this case, implying a nonlinear diffusion equation.

\subsection{Mesoscopic-Scale Rate Theory for Slow Defect Migration and Interaction \textmd{({\em defect reaction and migration}: $t>10$\,ns)}}
\label{sec2.3}

The formation, growth and dissolution of defect clusters such as voids, dislocation loops, etc., depend on the diffusion of point defects and their reaction with these defect clusters.\,\cite{book-gary} At the same time, they also depend on the concentration of point defects in the crystal. Since particle irradiation greatly raises the defect concentration above its thermal-equilibrium value, the diffusion coefficient can be enhanced. It can also be enhanced by the creation of new defect species.
\medskip

\subsubsection{Point-Defect Diffusion Equation}
\label{sec2.3.1}

By introducing the local coupling rates ${\cal R}({\bf r},\,t)$, $\Gamma_{\rm is}({\bf r},\,t)$ and $\Gamma_{\rm vs}({\bf r},\,t)$
for vacancy-interstitial recombination, interstitial-sink and vacancy-sink reaction rates,
we can write down the following two nonlinear rate-based diffusion equations for binary crystals

\[
\frac{\partial c_{\rm v}({\bf r},\,t)}{\partial t}=\nabla\cdot\left[D({\bf r},t\vert c_{\rm v})\nabla c_{\rm v}({\bf r},\,t)\right]-\nabla\cdot{\bf J}_{\rm v}({\bf r},t\vert c_{\rm v})
\]
\begin{equation}
+{\cal G}_0({\bf r},\,t)-{\cal R}({\bf r},\,t)\,c_{\rm i}({\bf r},\,t)\left[c_{\rm v}({\bf r},\,t)-c_{\rm v}^{\rm eq}({\bf r})\right]
-\Gamma_{\rm vs}({\bf r},\,t)\,c_{\rm s}({\bf r},\,t)\left[c_{\rm v}({\bf r},\,t)-c_{\rm v}^{\rm eq}({\bf r})\right]\ ,
\label{eqn-13}
\end{equation}

\[
\frac{\partial c_{\rm i}({\bf r},\,t)}{\partial t}=\nabla\cdot\left[D({\bf r},t\vert c_{\rm i})\nabla c_{\rm i}({\bf r},\,t)\right]-\nabla\cdot{\bf J}_{\rm i}({\bf r},t\vert c_{\rm i})
\]
\begin{equation}
+{\cal G}_0({\bf r},\,t)-{\cal R}({\bf r},\,t)\,c_{\rm i}({\bf r},\,t)\left[c_{\rm v}({\bf r},\,t)-c_{\rm v}^{\rm eq}({\bf r})\right]
-\Gamma_{\rm is}({\bf r},\,t)\,c_{\rm s}({\bf r},\,t)\,c_{\rm i}({\bf r},\,t)\ ,
\label{eqn-14}
\end{equation}
where we have neglected correlated diffusions and defect-defect interactions. Moreover, ${\bf J}_{\rm v}({\bf r},t\vert c_{\rm v})$ and ${\bf J}_{\rm i}({\bf r},t\vert c_{\rm i})$ in Eqs.\,(\ref{eqn-13}) and (\ref{eqn-14}) are the particle currents for vacancies and interstitials, and the master equation for determining the local sink concentration $c_{\rm s}({\bf r},\,t)$ will be given later.
\medskip

For simplicity, we consider a homogeneous system with volume ${\cal V}$ in the absence of vacancy and interstitial currents and assume all the rates are independent of time. We further neglect the small thermal-equilibrium vacancy concentration and write the defect generation rate as $G_0={\cal G}_0{\cal V}$. For this model system, we find that the evolution of $C_{\rm v}(t)$ and $C_{\rm i}(t)$ depends on the temperature and $C_{\rm s}$ and can be characterized in several regimes separated by different time scales $\tau$, including initial buildup without reaction, dominant vacancy-interstitial mutual recombination, and final vacancy and interstitial annihilation by sinks. As an example, we consider the case with low temperatures (much less than half of the melting temperature) and low sink densities. In the initial buildup regime-I with $0<t\leq \tau_1$, we have increasing $C^{({\rm I})}_{\rm v}(t)=C^{({\rm I})}_{\rm i}(t)=G_0t$ and $\tau_1=(G_0{\cal R})^{-1/2}$. In the next recombination regime-II with $\tau_1<t\leq\tau_2$, we get constant $C^{({\rm II})}_{\rm v}(t)=C_{\rm i}(t)=C^{({\rm II})}_1=(G_0/{\cal R})^{1/2}$ and $\tau_2=(\Gamma_{\rm is}C_{\rm s})^{-1}$. After this regime, we enter into the interstitial annihilation regime-III, where we find increasing $C^{({\rm III})}_{\rm v}(t)=(G_0\Gamma_{\rm is}C_{\rm s}t/{\cal R})^{1/2}$ and decreasing $C^{(\rm {III})}_{\rm i}(t)=(G_0/\Gamma_{\rm is}{\cal R}C_{\rm s}t)^{1/2}$ with $\tau_3=(\Gamma_{\rm vs}C_{\rm s})^{-1}$. Finally, in the vacancy annihilation regime-IV with $t>\tau_3$ (reaching a steady state), we arrive at constant $C^{({\rm IV})}_{\rm v}(t)=C_{2+}=(G_0\Gamma_{\rm is}/{\cal R}\Gamma_{\rm vs})^{1/2}$ and $C^{({\rm IV})}_{\rm i}(t)=C_{2-}=(G_0\Gamma_{\rm vs}/{\cal R}\Gamma_{\rm is})^{1/2}$. Physically, it is easy to understand that the first two regimes correspond to the `ultrafast' atomic-scale modeling, while the last two regimes are associated with the `slow' mesoscopic-scale modeling. By finding $C_{\rm v}$ and $C_{\rm i}$, we can calculate the radiation-enhanced diffusion coefficient $D_{\rm rad}=D_{\rm v}C_{\rm v}+D_{\rm i}C_{\rm i}$ due to large values of $C_{\rm v}$ and $C_{\rm i}$ in comparison with $C_{\rm v}^{\rm eq}$. In the steady state with high sink densities $C_{\rm s}$, we have $D_{\rm rad}=D_{\rm v}C_{\rm v}+D_{\rm i}C_{\rm v}\,(\Gamma_{\rm vs}/\Gamma_{\rm is})$, $C_{\rm v}=(G_0/\Gamma_{\rm vs}C_{\rm s})\,F(\eta)$, $\eta=4G_0{\cal R}/\Gamma_{\rm vs}\Gamma_{\rm is}C^2_{\rm s}$ and $F(\eta)=(2/\eta)\,(\sqrt{1+\eta}-1)$. If $\eta\rightarrow 0$, the defects are lost to sinks and none to recombination. If $\eta\gg 1$, on the other hand, the mutual recombination dominates the loss of defects.
\medskip

\subsubsection{Recombination and Sink Annihilation Rates}
\label{sec2.3.2}

In general, the reaction rate between species A and B can be expressed as $\Gamma_{AB}\,c_A\,c_B$ where $c_A$ and $c_B$ are the concentrations (particles/cm$^3$) and $\Gamma_{AB}$ (cm$^3$/s) is the rate constant.
\medskip

As an example, the recombination rate constant ${\cal R}({\bf r},\,t)$ in Eqs.\,(\ref{eqn-13}) and (\ref{eqn-14}) for vacancies and interstitials takes the form of

\begin{equation}
{\cal R}({\bf r},\,t)=\frac{z_{\rm iv}({\bf r})\Omega({\bf r})D_{\rm i}({\bf r},\,t)}{a^2_0({\bf r})}\ ,
\label{eqn-15}
\end{equation}
where $z_{\rm iv}({\bf r})$ (an integer) is the bias factor, depending on the crystal structure and species, $\Omega({\bf r})$ is the atomic volume, $a_0({\bf r})$ is the lattice constant and $D_{\rm i}({\bf r},\,t)$ is the mobil interstitial diffusion coefficient.
\medskip

In a similar way, the interstitial-sink $\Gamma_{\rm is}({\bf r},\,t)$ or the vacancy-sink $\Gamma_{\rm vs}({\bf r},\,t)$ annihilation rate constants
in Eqs.\,(\ref{eqn-13}) and (\ref{eqn-14}) are given by $\Gamma_{\rm \alpha s}({\bf r},\,t)\,c_{\alpha}({\bf r},\,t)\,c_{\rm s}({\bf r},\,t)
=\kappa^2_{\rm \alpha s}({\bf r},\,t)\,c_{\alpha}({\bf r},\,t)\,D_{\alpha}({\bf r},\,t)$, where $\alpha$ corresponds to mobil defect species and
$\kappa_{\rm \alpha s}({\bf r},\,t)$ (cm$^{-2}$) represents the sink strength given by

\begin{equation}
\kappa^2_{\rm \alpha s}({\bf r},\,t)=\frac{\Gamma_{\rm \alpha s}({\bf r},\,t)c_{\rm s}({\bf r},\,t)}{D_{\alpha}({\bf r},\,t)}\ .
\label{eqn-16}
\end{equation}
The sink strength measures the affinity of a sink for defects, which is independent of defect properties, and $\kappa^{-1}_{\rm \alpha s}({\bf r},\,t)$
corresponds to the mean distance for a traveling defect in the crystal before it is trapped by sinks.

\subsubsection{Point-Defect Interaction Rates}
\label{sec2.3.3}

In the absence of the macroscopic-scale gradient of defect concentration, the reaction between defects and sinks is reaction-rate-controlled. According to Eq.\,(\ref{eqn-15}), the defect-void interaction can be described by the rate constants $\Gamma_{\rm \{i,v\}V}({\bf r},\,t)$ given by

\begin{equation}
\Gamma_{\rm \{i,v\}V}({\bf r},\,t)=\sum_{n=2}^{\infty}\,\frac{4\pi R^2({\bf r},t\vert n)D_{\rm i,v}({\bf r},\,t)}{a_0({\bf r})}
=\sum_{n=2}^{\infty}\,\frac{\kappa^2_{\rm V}({\bf r},t\vert n)D_{\rm i,v}({\bf r},\,t)}{c_{\rm V}({\bf r},t\vert n)}\ ,
\label{eqn-17}
\end{equation}
where $z_{\rm \{i,v\}V}=4\pi R^2/a_0^2$, $\Omega\sim a_0^3$, $R({\bf r},t\vert n)$ represents the radius of a void sphere involving $n$ vacancies. The void strength is given by $\kappa^2_{\rm V}({\bf r},t\vert n)=4\pi R^2({\bf r},t\vert n)c_{\rm V}({\bf r},t\vert n)/a_0({\bf r})$, where $c_{\rm V}({\bf r},t\vert n)$ is the concentration of voids containing $n$ vacancies in the crystal.
\medskip

Similarly, for the defect-dislocation interaction, we have the rate constants $\Gamma_{\rm \{i,v\}d}({\bf r},\,t)$ (in units of cm$^2$/s) given by

\begin{equation}
\Gamma_{\rm \{i,v\}d}({\bf r},\,t)=z_{\rm \{i,v\}d}({\bf r})D_{\rm i,v}({\bf r},\,t)=\frac{\kappa^2_{\rm \{i,v\}d}({\bf r},\,t)D_{\rm i,v}({\bf r},\,t)}
{\rho_{\rm d}({\bf r},\,t)}\ ,
\label{eqn-18}
\end{equation}
where we replace $\Omega$ in Eq.\,(\ref{eqn-15}) by an atomic area ($\sim a_0^2$), $z_{\rm id}({\bf r})\neq z_{\rm vd}({\bf r})$, $\rho_{\rm d}({\bf r},\,t)$ is the dislocation areal density and the dislocation capture rate per unit volume is

\begin{equation}
{\cal Q}_{\rm \{i,v\}d}({\bf r},\,t)=z_{\rm \{i,v\}d}({\bf r})D_{\rm i,v}({\bf r},\,t)({\bf r})\rho_{\rm d}({\bf r},\,t)c_{\rm i,v}({\bf r},\,t)\ .
\label{eqn-19}
\end{equation}
\medskip

Reactions driven by defect concentration gradients are diffusion limited instead of reaction-rate limited as discussed above. In this case, we have to solve the
diffusion term $\nabla\cdot\left[D_{\rm i,v}({\bf r},\,t)\nabla c_{\rm i,v}({\bf r},\,t)\right]$ with the generation term ${\cal G}_0({\bf r},\,t)$ for spherical (voids) or cylindrical (dislocation lines) coordinates.
\medskip

For the defect-void interaction, we get $\displaystyle{\Gamma_{\rm \{i,v\}V}({\bf r},\,t)=\sum_{n=2}^{\infty}\,\Gamma_{\rm \{i,v\}V}({\bf r},t\vert n)}$,
$\Gamma_{\rm \{i,v\}V}({\bf r},t\vert n)=4\pi R({\bf r},t\vert n)\,D_{\rm i,v}({\bf r},\,t)$ and $\kappa^2_{\rm V}({\bf r},t\vert n)=4\pi R({\bf r},t\vert n)\,c_{\rm V}({\bf r},t\vert n)$. For the defect-dislocation line interaction, on the other hand, we have $\Gamma_{\rm \{i,v\}d}({\bf r},\,t)=2\pi D_{\rm i,v}({\bf r},\,t)/\ln[R_0/R_{\rm \{i,v\}d}({\bf r},\,t)]$ and $\kappa^2_{\rm \{i,v\}d}({\bf r},\,t)=2\pi\rho_{\rm d}({\bf r},\,t)/\ln[R_0/R_{\rm \{i,v\}d}({\bf r},\,t)]$, where $R_0$ is the absorption radius of a sink, $R_{\rm id}({\bf r},\,t)$ and $R_{\rm vd}({\bf r},\,t)$ are the sink capture radii for interstitials and vacancies, respectively, with $R_{\rm id}\gg R_{\rm vd}$.
\medskip

In the presence of grain boundaries, we obtain the grain-boundary sink strength

\[
\kappa^2_{\rm \{i,v\}gb}({\bf r},\,t)=4\pi R_{\rm gb}({\bf r})c_{\rm gb}({\bf r})
\]
\begin{equation}
\times\left\{\frac{\kappa_{\rm i,v}({\bf r},\,t)R_{\rm gb}({\bf r})\cosh[\kappa_{\rm i,v}({\bf r},\,t)R_{\rm gb}({\bf r})]-\sinh[\kappa_{\rm i,v}({\bf r},\,t)R_{\rm gb}({\bf r})]}{\sinh[\kappa_{\rm i,v}({\bf r},\,t)R_{\rm gb}({\bf r})]-\kappa_{\rm i,v}({\bf r},\,t)R_{\rm gb}({\bf r})}\right\}\ ,
\label{eqn-20}
\end{equation}
where $R_{\rm gb}({\bf r})$ is the radius of a spherical grain, $c_{\rm gb}({\bf r})$ is the grain concentration and $\kappa_{\rm i,v}({\bf r},\,t)$ is the
sink strength for the grain interior for interstitlals or vacancies due to dislocations and voids. Moreover, its rate constant is $\Gamma_{\rm \{i,v\}gb}({\bf r},\,t)=\kappa^2_{\rm \{i,v\}gb}({\bf r},\,t)D_{\rm i,v}({\bf r},\,t)/c_{\rm gb}({\bf r})$.

\subsubsection{Radiation-Induced Segregation}
\label{sec2.3.4}

For a binary A-B alloy (or donor and acceptor randomly-doped semiconductors), in the absence of sinks, the diffusion equations for vacancies, interstitials and atoms A and B are

\[
\frac{\partial c_{\rm v}({\bf r},\,t)}{\partial t}=-\nabla\cdot{\bf J}_{\rm v}({\bf r},\,t)+{\cal G}_0({\bf r},\,t)-{\cal R}({\bf r},\,t)c_{\rm i}({\bf r},\,t)c_{\rm v}({\bf r},\,t)
\]
\[
=\nabla\left\{-\left[d_{\rm Av}({\bf r})-d_{\rm Bv}({\bf r})\right]\chi({\bf r},\,t)\Omega({\bf r})c_{\rm v}({\bf r},\,t)\nabla c_{\rm A}({\bf r},\,t)+D_{\rm v}({\bf r},\,t)
\nabla c_{\rm v}({\bf r},\,t)\right\}
\]
\begin{equation}
+{\cal G}_0({\bf r},\,t)-{\cal R}({\bf r},\,t)c_{\rm i}({\bf r},\,t)c_{\rm v}({\bf r},\,t)\ ,
\label{eqn-21}
\end{equation}

\[
\frac{\partial c_{\rm i}({\bf r},\,t)}{\partial t}=-\nabla\cdot{\bf J}_{\rm i}({\bf r},\,t)+{\cal G}_0({\bf r},\,t)-{\cal R}({\bf r},\,t)c_{\rm i}({\bf r},\,t)c_{\rm v}({\bf r},\,t)
\]
\[
=\nabla\left\{\left[d_{\rm Ai}({\bf r})-d_{\rm Bi}({\bf r})\right]\chi({\bf r},\,t)\Omega({\bf r})c_{\rm i}({\bf r},\,t)\nabla c_{\rm A}({\bf r},\,t)+D_{\rm i}({\bf r},\,t)
\nabla c_{\rm i}({\bf r},\,t)\right\}
\]
\begin{equation}
+{\cal G}_0({\bf r},\,t)-{\cal R}({\bf r},\,t)c_{\rm i}({\bf r},\,t)c_{\rm v}({\bf r},\,t)\ ,
\label{eqn-22}
\end{equation}

\[
\frac{\partial c_{\rm A}({\bf r},\,t)}{\partial t}=-\nabla\cdot{\bf J}_{\rm A}({\bf r},\,t)=\nabla\left\{D_{\rm A}({\bf r})\chi({\bf r},\,t)\nabla c_{\rm A}({\bf r},\,t)\right.
\]
\begin{equation}
\left.+\Omega({\bf r})c_{\rm A}({\bf r},\,t)\left[d_{\rm Ai}({\bf r})\nabla c_{\rm i}({\bf r},\,t)-d_{\rm Av}({\bf r})\nabla c_{\rm v}({\bf r},\,t)\right]\right\}\ ,
\label{eqn-23}
\end{equation}
where $d_{\rm \{A,B\}\{i,v\}}({\bf r})=\lambda^2_{\rm i,v}({\bf r})z_{\rm i,v}({\bf r})\omega_{\rm \{A,B\}\{i,v\}}({\bf r})$ are the diffusivity coefficients and the dimensionless $\chi({\bf r},\,t)$ is the thermodynamic factor connecting the chemical-potential gradient to the concentration gradient. In addition, we have $c_{\rm B}({\bf r},\,t)=\Omega^{-1}({\bf r})-c_{\rm A}({\bf r},\,t)$ when small defect concentrations are neglected.
\medskip

By requiring ${\bf J}_{\rm A}={\bf J}_{\rm B}=0$ and ${\bf J}_{\rm i}={\bf J}_{\rm v}$ for steady state and neglecting ${\cal G}_0({\bf r},\,t)$ and ${\cal R}({\bf r},\,t)$ in Eqs.\,(\ref{eqn-21})-(\ref{eqn-23}), we get

\begin{equation}
\nabla c_{\rm A}=-\nabla c_{\rm B}=\frac{N_{\rm A}N_{\rm B}d_{\rm Bi}d_{\rm Ai}}{\chi(d_{\rm Bi}N_BD_{\rm A}+d_{\rm Ai}N_{\rm A}D_{\rm B})}
\left(\frac{d_{\rm Av}}{d_{\rm Bv}}-\frac{d_{\rm Ai}}{d_{\rm Bi}}\right)\nabla c_{\rm V}\ ,
\label{eqn-24}
\end{equation}
where $N_{\rm A,B}=c_{\rm A,B}\Omega$ and the direction of $\nabla c_{\rm A}$ can be either parallel or anti-parallel to $\nabla c_{\rm v}$. Additionally, the undersized (oversized) solutes bounded to interstitials will be concentrated (depleted) around sinks to create a concentration gradient after their redistribution.
\medskip

On the other hand, the oversized or undersized solutes with respect to the lattice atoms can act as traps for vacancies and interstitials, including release of defects from traps, recombination with trapped defects, trapping of free point defects and loss to internal sinks. This is further supplemented by three rate equations for trap and trapped defect concentrations.

\section{Sinks Dynamics}
\label{sec3}

The growth of dislocation loops and spherical voids is determined by solving the point-defect balance equations without diffusion terms. Since the defect concentration is still changing with time due to the time-dependent radiation source (or defect production rate), only quasi-steady state can be defined for short periods of time. Physically, the quasi-steady state is related to the fact that the change in sink strength due to microstructure evolution is slow  compared to
the response time of the defect population.

\subsection{Crystal Elasticity}
\label{sec3.1}

For a given displacement vector $\boldsymbol{u}({\bf r})=[u_1({\bf r}),\,u_2({\bf r}),\,u_3({\bf r})]$, the symmetric strain tensor $\underline{\boldsymbol{\epsilon}}=[\epsilon_{ij}({\bf r})]$ is defined as

\begin{equation}
\epsilon_{ij}({\bf r})=\frac{1}{2}\left[\frac{\partial u_i({\bf r})}{\partial x_j}+\frac{\partial u_j({\bf r})}{\partial x_i}\right]=\epsilon_{j\,i}({\bf r})\ ,
\label{eqn-59}
\end{equation}
where ${\bf r}=(x_1,\,x_2,\,x_3)$ is the position vector in space. The elastic force ${\bf F}({\bf r})=[F_1({\bf r}),\,F_2({\bf r}),\,F_3({\bf r})]$ per unit volume is given by

\begin{equation}
F_i({\bf r})=\sum\limits_{j=1}^{3}\,\frac{\partial\sigma_{ij}({\bf r})}{\partial x_j}\ ,
\label{eqn-25}
\end{equation}
where the stress tensor $\underline{\boldsymbol{\sigma}}=[\sigma_{ij}({\bf r})]$ is related to the strain tensor $\underline{\boldsymbol{\epsilon}}=[\epsilon_{ij}({\bf r})]$ by Hooke's law.

\subsection{Planar Biased Dislocation-Loop Growth}
\label{sec3.2}

By defining the dislocation line direction ${\bf s}$ and the Burgers vector\,\cite{book-gary} ${\bf b}$ for edge (${\bf b}\perp{\bf s}$) or skew (${\bf b}\|{\bf s}$) dislocations, the Peach-Koehler equation\,\cite{book-gary} gives us the force ${\bf f}$ per length as

\begin{equation}
{\bf f}={\bf b}^{\rm T}\cdot(\underline{\boldsymbol{\sigma}}\times{\bf s})\ ,
\label{eqn-26}
\end{equation}
where $\underline{\boldsymbol{\sigma}}$ is the stress tensor. The force ${\bf f}$ along the ${\bf b}$ direction is the glide force, while ${\bf f}$ perpendicular to both ${\bf s}$ and ${\bf b}$ directions is called the climb force. The Peach-Koehler equation can be used for calculating the interaction between dislocations, where ${\bf b}$ and ${\bf s}$ are assigned to the dislocation-2 while $\underline{\sigma}$ is for the dislocation-1. For the edge dislocation, we have five non-zero stress tensor elements $\sigma_{xx}$, $\sigma_{yy}$, $\sigma_{zz}$ and $\sigma_{xy}=\sigma_{yx}$, while for the skew dislocation, we only have four non-zero stress tensor elements $\sigma_{yz}=\sigma_{zy}$ and $\sigma_{xz}=\sigma_{zx}$.
\medskip

Besides the dislocation lines, there also exists Frank loops. For example, the close-packed {\em fcc} lattice follows the stacking sequence $ABCABCABC\cdots$, where $A$, $B$ and $C$ correspond to different planes of atoms. It can be modified to $ABCAB/ABC\cdots$, where ``$/$'' denotes the intrinsic single fault or missing plane of atoms. It can also be modified to $ABCAB/A/CABC\cdots$, where a plane of atoms or the extrinsic double fault is inserted.
\medskip

Interstitial condensation can occur around the edges of the depleted zone. A cluster of point defects can be a line, a disc or a void. The formation of the perfect or faulted loops of interstitials competes with the formation of the voids of vacancies, which are also affected by the irradiation temperature.
\medskip

The nucleation of loops is a clustering process that results in a critical size embryo for further growing. As an example, by denoting the number of clusters  consisting of $j$ vacancies as $\rho_{\rm v}(j)$, the {\em master equations} for $\rho_{\rm v}(j)$ are

\[
\frac{\partial\rho_{\rm v}(j)}{\partial t}={\cal G}_{0}(j)-\sum\limits_{n=1}^{\infty}\left[\beta_{\rm v_n}(j)+\beta_{\rm i_n}(j)\right]\rho_{\rm v}(j)
-\sum\limits_{n=1}^{j}\,\alpha_{\rm v_n}(j)\rho_{\rm v}(j)
\]
\begin{equation}
+\sum\limits_{n=1}^{j-1}\,\beta_{\rm v_n}(j-n)\rho_{\rm v}(j-n)
+\sum\limits_{n=1}^{\infty}\,\beta_{\rm i_n}(j-n)\rho_{\rm v}(j+n)
+\sum\limits_{n=1}^{\infty}\,\alpha_{\rm v_n}(j+n)\rho_{\rm v}(j+n)\ ,
\label{eqn-27}
\end{equation}
where $\beta_{\rm v_n}$ and $\beta_{\rm i_n}$ are the capture rates of migrating vacancy (v) or interstitila (i) clusters of size $n$ by a cluster of size $j$,
and $\alpha_{\rm v_n}(j)$ is the emission rate for the new vacancy cluster of size $n$ by a cluster of size $j$. In Eq.\,(\ref{eqn-27}), the first term is the direct production of a cluster of size $j$, while the second term is the loss of clusters from size $j$ due to absorption of a cluster of size $n$. The third term is the loss of a cluster of size $j$ due to emission of a cluster of size $n$. The fourth and fifth terms in Eq.\,(\ref{eqn-27}) are the addition of clusters to the cluster of size $j$ due to absorption of vacancy clusters by a smaller cluster and absorption of interstitial clusters by a larger cluster, and the last term is the addition of clusters to the cluster of size $j$ due to loss of vacancy clusters by a larger cluster.
\medskip

Since the dominant contribution for cluster reactions is with point defects (i.e., cluster of size $j=1$), for both vacancies and interstitials, Eq.\,(\ref{eqn-27}) with $j\geq 2$ can be simplified to

\[
\frac{\partial\rho(j;\,t)}{\partial t}={\cal G}_{0}(j;\,t)+\beta(j-1,\,j)\rho(j-1;\, t)
+\alpha(j+1,\,j)\rho(j+1;\,t)
\]
\begin{equation}
-\left[\beta(j,\,j+1)+\alpha(j,\,j-1)\right]\rho(j;\,t)\ .
\label{eqn-28}
\end{equation}
If the cluster size index $j$ can be treated as a continuous variable $\xi$, Eq.\,(\ref{eqn-28}) reduces to a {\em Fokker-Planck equation} as\,\cite{woo}

\begin{equation}
\frac{\partial\rho(\xi;\,t)}{\partial t}={\cal G}_{0}(\xi;\,t)-\frac{\partial}{\partial\xi}
\left\{{\cal F}(\xi;\,t)\rho(\xi;\,t)-\frac{\partial}{\partial\xi}\left[{\cal D}(\xi;\,t)\rho(\xi;\,t)\right]\right\}\ ,
\label{eqn-29}
\end{equation}
where the second term in the equation represents the drift in size space to larger vacancy clusters, and the last term is the diffusion in size space to a broader cluster size distribution. Additionally, we have defined two coefficients in Eq.\,(\ref{eqn-29}), i.e., ${\cal F}(\xi;\,t)=[z_{\rm v}({\bf r})D_{\rm v}({\bf r},\,t)c_{\rm v}({\bf r},\,t)-z_{\rm i}({\bf r})D_{\rm i}({\bf r},\,t)c_{\rm i}({\bf r},\,t)]$, which determines the direction of the drift, and the positive ${\cal D}(\xi;\,t)=(1/2)[z_{\rm v}({\bf r})D_{\rm v}({\bf r},\,t)c_{\rm v}({\bf r},\,t)+z_{\rm i}({\bf r})D_{\rm i}({\bf r},\,t)c_{\rm i}({\bf r},\,t)]$.
\medskip

According to Eq.\,(\ref{eqn-28}), for the dislocation loop growth, we find the evolution of the number density $\rho_{\rm il}(j,\,t)$ for the interstitial loop of size $j$ satisfies

\[
\left.\frac{\partial\rho_{\rm il}(j;\,t)}{\partial t}={\cal G}_{0}(j;\,t)\right|_{j\geq 4}+\left[\beta_{\rm v}(j+1)+\alpha_{\rm i}(j+1)\right]\rho_{\rm il}(j+1;\, t)
+\beta_{\rm i}(j-1)\rho_{\rm il}(j-1;\,t)
\]
\begin{equation}
-\left[\beta_{\rm v}(j)+\beta_{\rm i}(j)+\alpha_{\rm i}(j)\right]\rho_{\rm il}(j;\,t)\ ,
\label{eqn-80}
\end{equation}
where

\begin{equation}
\beta_{\rm i,v}(j)=2\pi\,r(j)z_{\rm c}(j)D_{\rm i,v}({\bf r},\,t)C_{\rm i,v}({\bf r},\,t)\ ,
\label{eqn-30}
\end{equation}

\begin{equation}
\alpha_{\rm i,v}(j)=2\pi\,r(j)z_{\rm c}(j)\frac{D_{\rm i,v}({\bf r},\,t)}{\Omega}\exp\left[-\frac{E_{\rm b,\{i,v\}}(j)}{k_{\rm B}T}\right]\ ,
\label{eqn-31}
\end{equation}
$r(j)$ and $z_{\rm c}$ are the radius and bias factor of an interstitial loop of size $j$, and $E_{\rm b}(j)$ is the binding energy for a cluster of $j$ defects.
\medskip

The saturation of the dislocation density $\rho_{\rm d}(t)$ in quasi-steady state was found experimentally to be due to a recovery process at high temperatures,\,\cite{book-gary} given by

\begin{equation}
\frac{d\rho_{\rm d}(t)}{dt}=B\rho_{\rm d}(t)-A\rho_{\rm d}^2(t)\ ,
\label{eqn-32}
\end{equation}
where $B$ and $A$ are constants. This gives the steady-state solution $\rho_{\rm d}^{\rm s}=B/A$, and the time-dependent solution is found to be

\begin{equation}
\frac{\rho_{\rm d}(t)}{\rho_{\rm d}^{\rm s}}=\left[\frac{1-e^{-x}+\sqrt{\rho_{\rm d}^0/\rho_{\rm d}^{\rm s}}(1+e^{-x})}{1+e^{-x}+\sqrt{\rho_{\rm d}^0/\rho_{\rm d}^{\rm s}}(1-e^{-x})}\right]^2\ ,
\label{eqn-33}
\end{equation}
where $\rho_{\rm d}^0$ is the initial value and $x(t)=A\sqrt{\rho_{\rm d}^{\rm s}}\,t$.

\subsection{Spherical Neutral Void Growth}
\label{sec3.3}

Not all the defects generated by radiation-induced atom displacement are point defects. Some of the defects form clusters, and the vacancy clusters, which are usually not stable and immobile, may further grow to form voids. For a small number of vacancies, the spherical void is favorable, while for a large number of vacancies, the planar loop is a more stable configuration.
\medskip

The dynamics for void growth is very similar to that for dislocation-loop growth. The net absorption rate of vacancies by a spherical void is the difference of absorption rates of vacancies and interstitials, i.e., $A^{\rm V}_{\rm net}=4\pi R_{\rm V}\{D_{\rm v}[V_{\rm v}-C_{\rm v}(R)]-D_{\rm i}C_{\rm i}\}$. Therefore, the equation for the growth of a spherical void of radius $R_{\rm V}(t)$ (or volume) in quasi-steady state is

\begin{equation}
\frac{dR_{\rm V}(t)}{dt}=\frac{\Omega}{R_{\rm V}(t)}\left\{D_{\rm v}({\bf r},\,t)\left[C_{\rm v}({\bf r},\,t)-C_{\rm v}(R_{\rm V})\right]-D_{\rm i}({\bf r},\,t)C_{\rm i}({\bf r},\,t)\right\}\ ,
\label{eqn-34}
\end{equation}
where $C_{\rm v}(R_{\rm V})$ is the vacancy concentration at the void surface.
\medskip

From the balance equation, we get the concentrations of point vacancies and interstitials as follows:

\begin{equation}
C_{\rm i,v}({\bf r},\,t)=\frac{D_{\rm i,v}({\bf r},\,t)\kappa^2_{\rm i,v}({\bf r},\,t)}{2{\cal R}({\bf r},\,t)}\left[\sqrt{\eta+1}-1\right]\ ,
\label{eqn-35}
\end{equation}
where $\kappa_{\rm i,v}^2({\bf r},\,t)=z_{\rm i,v}({\bf r})\rho_{\rm d}({\bf r},\,t)+4\pi R_{\rm V}({\bf r},\,t)c_{\rm V}({\bf r},\,t)$, $\eta=4{\cal R}{\cal G}_0/D_{\rm i}D_{\rm v}\kappa_{\rm v}^2\kappa_{\rm i}^2$. Inserting Eq.\,(\ref{eqn-35}) into Eq.\,(\ref{eqn-34}), we obtain

\begin{equation}
\frac{dR_{\rm V}(t)}{dt}=\frac{dR_0(t)}{dt}\left\{\frac{2}{\eta}\left[\sqrt{\eta+1}-1\right]\right\}-\frac{dR_{\rm th}(t)}{dt}\ ,
\label{eqn-36}
\end{equation}
where $dR_0(t)/dt$ is proportional to $(z_{\rm i}-z_{\rm v})\,{\cal G}_0$ and independent of temperature, while the second negative term represents the thermal emission of defects from sinks and strongly depends on temperature (proportional to $D_{\rm v}C_{\rm v}^{\rm eq}$). The thermal emission of defects from sinks can be calculated as

\begin{equation}
\frac{dR_{\rm th}(t)}{dt}=\frac{D_{\rm v}C_{\rm v}^{\rm th}\Omega^2z_{\rm v}\rho_{\rm d}(2\gamma/R_{\rm V}-\sigma-P)}{R_{\rm V}k_{\rm B}T(z_{\rm v}\rho_{\rm d}+4\pi R_{\rm V}c_{\rm V})}\ ,
\label{eqn-39}
\end{equation}
where $\sigma$ is the hydrostatic stress, $P$ is the gas pressure in the void, and $\gamma$ represents the surface tension. In the absence of surface tension and gas pressure, we have $dR_{\rm th}(t)/dt>0$, indicating a shrinkage of the void in Eq.\,(\ref{eqn-36}).
\medskip

At low temperatures, both $D_{\rm v}$ and $C_{\rm v}^{\rm eq}$ are small. This leads to large $\eta$ and decreased $dR_{\rm V}(t)/dt$, so that vacancy concentration is built up and vacancies and interstitials are lost to recombination. At high temperatures, on the other hand, the thermal emission of vacancies becomes dominant and compensates the void growth. As a result, we generally expect a maximized $dR_{\rm V}(t)/dt$ at intermediate temperature. The maximum value reduces with increasing ${\cal G}_0$ and the corresponding peak temperature shifts up with ${\cal G}_0$. For a constant radiation dose, the temperature shift $\Delta T$ is determined by

\begin{equation}
\frac{\Delta T}{T_1}=\frac{T_2-T_1}{T_1}=\frac{(k_{\rm B}T_1/E_m^{\rm v})\ln({\cal G}_{02}/{\cal G}_{01})}{1-(k_{\rm B}T_1/E_m^{\rm v})\ln({\cal G}_{02}/{\cal G}_{01})}\ ,
\label{eqn-38}
\end{equation}
where $E_m^{\rm v}$ is the migration energy in the vacancy diffusion coefficient $D_{\rm v}(T)=D_0\exp(-E_m^{\rm v}/k_{\rm B}T)$.
\medskip

Neglecting the thermal emission in Eq.\,(\ref{eqn-36}), we get from Eq.\,(\ref{eqn-34})

\begin{equation}
\frac{dR_{\rm V}}{dt}\approx\frac{\Omega D_{\rm v}D_{\rm i}}{2R_{\rm V}{\cal R}}\left[\sqrt{1+\frac{4{\cal R}{\cal G}_0}
{D_{\rm i}D_{\rm v}\kappa_{\rm v}^2\kappa_{\rm i}^2}}-1\right]\rho_{\rm d}\left(z_{\rm i}^{\rm d}z_{\rm v}-z_{\rm v}^{\rm d}z_{\rm i}\right)\ ,
\label{eqn-37}
\end{equation}
where the sign of the bias of dislocation $(z_{\rm i}^{\rm d}z_{\rm v}-z_{\rm v}^{\rm d}z_{\rm i})$ for vacancies and interstitials\,\cite{book-gary} determines the occurrence of either growth ($dR_{\rm V}/dt>0$) or shrinkage ($dR_{\rm V}/dt<0$). In the sink-dominant process with $(4{\cal R}{\cal G}_0/D_{\rm i}D_{\rm v}\kappa_{\rm v}^2\kappa_{\rm i}^2)\ll 1$, we find $dR_{\rm V}/dt\propto{\cal G}_0$. If recombination dominates, i.e., $(4{\cal R}{\cal G}_0/D_{\rm i}D_{\rm v}\kappa_{\rm v}^2\kappa_{\rm i}^2)\gg 1$, we obtain $dR_{\rm V}/dt\propto\sqrt{{\cal G}_0}$. Moreover, the rate for the void growth reaches a maximum when the dislocation sink strength $z^{\rm d}_{\rm v}\rho_{\rm d}$ equals to the void sink strength $4\pi R_{\rm V}c_{\rm V}z_{\rm v}$. If $z^{\rm d}_{\rm v}\rho_{\rm d}>4\pi R_{\rm V}c_{\rm V}z_{\rm v}$, we find that the void concentration $c_{\rm V}$ is low. On the other hand, we find a high void concentration for $z^{\rm d}_{\rm v}\rho_{\rm d}<4\pi R_{\rm V}c_{\rm V}z_{\rm v}$.

\section{Radiation Degradation of Electronic Devices}
\label{sec4}

\begin{figure}[p]
\centering
\epsfig{file=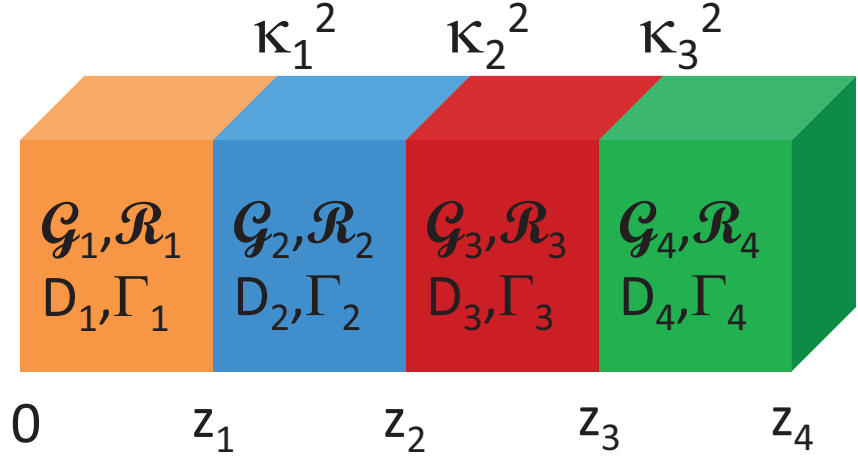,width=0.7\textwidth}
\caption{Layered structure materials with radiation parameters ${\cal G}_j$, ${\cal R}_j$, $D_j$ and $\Gamma_j$ ($j=1,\,2,\,3,\,4$) for generation, recombination, diffusion coefficient and bulk-sink annihilation, respectively. In addition, $\kappa^2_i$ for $i=1,\,2,\,3$ represents the interface-sink strength. Particles are incident from the front surface at $z=0$ and exit from the back surface at $z=z_4$.}
\label{f3}
\end{figure}

Let us consider a commonly used layered-structure material,\,\cite{qw} as shown in Fig.\,\ref{f3}. Each material layer is characterized by the radiation parameters ${\cal G}_j$, ${\cal R}_j$, $D_j$ and $\Gamma_j$ with $j=1,\,2,\,3,\,4$ for generation and recombination rates, diffusion coefficient and bulk-sink annihilation, which will be employed to model from an ultrafast atomic-scale up to $100$\,ns. The calculated non-steady state defect distribution in each layer will be used for initial conditions in a slow mesoscopic-scale diffusion and annihilation model in order to calculate the steady-state spatial distribution of defects in the whole layered structure. In  modeling the mesoscopic-scale, the interface-sink strengths $\kappa^2_i$ with $i=1,\,2,\,3$ will also be considered. Once the steady-state distribution of point defects, denoted as $\rho_{\rm d}(z)$, is obtained for the whole layered structure, they will be fed into the follow-up calculations for radiation degradation in electronic devices, as described below.
\medskip

The band structure of a crystal largely determines the properties of electrons,\,\cite{book-huang} such as effective mass, bandgap energy, density of states, plasma frequency and absorption coefficient. These electron properties are a result of unique crystal potential from all lattice atoms, instead of properties of an individual lattice atom. On the other hand, the radiation-induced displacements of lattice atoms are determined not only by the intrinsic properties, such as mass of the atoms, but also by the extrinsic conditions,\,\cite{book-gary} such as kinetic energy of incident particles and lattice temperature.

\subsection{Steady-State Defect Distributions}
\label{sec4.1}

For the reaction rate control system shown in Fig.\,\ref{f3}, by generalizing Eqs.\,(\ref{eqn-13}) and (\ref{eqn-14}), we write down the diffusion equations for point vacancies and interstitials as

\[
\frac{\partial c^{j}_{\rm v}(z,\,t)}{\partial t}=D^{j}_{\rm v}\,\frac{\partial^2 c^{j}_{\rm v}(z,\,t)}{\partial z^2}+{\cal G}^{j}_0(t)-\left[\frac{z^{j}_{\rm iv}\Omega_{j}D^{j}_{\rm i}}{(a_0^{j})^2}\right]c^{j}_{\rm i}(z,\,t)\left[c^{j}_{\rm v}(z,\,t)-c_{\rm v}^{\rm eq}(j)\right]
\]
\[
-\sum\limits_{n=2}^{\infty}\left\{\frac{4\pi[R_{\rm V}^{j}(t\vert n)]^2D^{j}_{\rm v}}{a^{j}_0}\right\}c^{j}_{\rm V}(z,t\vert n)\left[c^{j}_{\rm v}(z,\,t)-c_{\rm v}^{\rm eq}(j)\right]
\]
\[
-\sum\limits_{n=2}^{\infty}\left[z^{j+1}_{\rm vd}\,\rho_{\rm d}^{j+1}(t\vert n)\,D^{j+1}_{\rm v}\right]\left[c^{j+1}_{\rm v}(z,\,t)-c_{\rm v}^{\rm eq}(j+1)\right]a_0^{j+1}\,\delta(z-z_{j+1})
\]
\begin{equation}
-\sum\limits_{n=2}^{\infty}\left[z^{j}_{\rm vd}\,\rho_{\rm d}^{j}(t\vert n)\,D^{j}_{\rm v}\right]\left[c^{j}_{\rm v}(z,\,t)-c_{\rm v}^{\rm eq}(j)\right]a_0^{j}\,\delta(z-z_{j})\ ,
\label{eqn-40}
\end{equation}

\[
\frac{\partial c^{j}_{\rm i}(z,\,t)}{\partial t}=D^{j}_{\rm i}\,\frac{\partial^2 c^{j}_{\rm i}(z,\,t)}{\partial z^2}+{\cal G}^{j}_0(t)-\left[\frac{z^{j}_{\rm iv}\Omega_{j}D^{j}_{\rm i}}{(a_0^{j})^2}\right]c^{j}_{\rm i}(z,\,t)\left[c^{j}_{\rm v}(z,\,t)-c_{\rm v}^{\rm eq}(j)\right]
\]
\[
-\sum\limits_{n=2}^{\infty}\left\{\frac{4\pi[R_{\rm V}^{j}(t\vert n)]^2D^{j}_{\rm i}}{a^{j}_0}\right\}c^{j}_{\rm V}(z,t\vert n)\,c^{j}_{\rm i}(z,\,t)
\]
\[
-\sum\limits_{n=2}^{\infty}\left[z^{j+1}_{\rm id}\,\rho_{\rm d}^{j+1}(t\vert n)\,D^{j+1}_{\rm i}\right]c^{j+1}_{\rm i}(z,\,t)\,a_0^{j+1}\,\delta(z-z_{j+1})
\]
\begin{equation}
-\sum\limits_{n=2}^{\infty}\left[z^{j}_{\rm vd}\,\rho_{\rm d}^{j}(t\vert n)\,D^{j}_{\rm i}\right]c^{j}_{\rm i}(z,\,t)\,a_0^{j}\,\delta(z-z_{j})\ ,
\label{eqn-41}
\end{equation}
where the integer $j$ is the layer index, $z_{j-1}$ and $z_{j}$ represent the left and right interface positions of the $j$th layer, respectively. In Eqs.\,(\ref{eqn-40}) and (\ref{eqn-41}), we used the facts that $\Gamma_{\rm i,v}=z_{\rm i,v}\Omega D_{\rm i,v}/a_0^2$, $\kappa^2_{V}=4\pi R_{\rm V}^2c_{\rm V}/a_0$ and $\kappa^2_{\rm \{i,v\}d}=z_{\rm \{i,v\}d}\rho_{\rm d}$ for a reaction rate control system.
\medskip

The diffusion coefficients $D^j_{\rm i,v}$ for point vacancies and interstitials are calculated as

\begin{equation}
D^j_{\rm i,v}(T)=f^{j}_{\rm i,v}\eta_{j}\,(a_0^{j})^2\nu_j\,\exp\left[\frac{\Delta S^{\rm i,v}_m(j)}{k_{\rm B}}\right]\,\exp\left[-\frac{E_m^{\rm i,v}(j)}{k_{\rm B}T}\right]
=f^{j}_{\rm i,v}D_0\,\exp\left[-\frac{E_m^{\rm i,v}(j)}{k_{\rm B}T}\right]\ ,
\label{eqn-81}
\end{equation}
where $f^{j}_{\rm i,v}<1$ is the diffusion correlation factor, $\eta_{j}$ is the structural factor relating to the jump distance and number of nearest neighbors, $\nu_j$ is the jump frequency on the order of the Debye frequency, $\Delta S^{\rm i,v}_m(j)$ is the change in entropy due to vibrational defect disorder, and $E_m^{\rm i,v}(j)$ is the point-defect migration energy.
\medskip

The radius $R_{\rm V}^{j}(t\vert n)$ of the spherical void of size $n$ (in unit of lattice constant) introduced in Eqs.\,(\ref{eqn-40}) and (\ref{eqn-41}) is determined from the following void growth equation:

\begin{equation}
\frac{dR^{j}_{\rm V}(t\vert n)}{dt}=\frac{\Omega_{j}}{R^{j}_{\rm V}(t\vert n)}\left\{D^{j}_{\rm v}\left[c^{j}_{\rm v}(z,\,t)-c^{j}_{\rm v}(t\vert R_{\rm V})\right]-D^{j}_{\rm i}\,c^{j}_{\rm i}(z,\,t)\right\}\ ,
\label{eqn-42}
\end{equation}
where $c^{j}_{\rm v}(t\vert R_{\rm V})$ is the vacancy concentration at the void surface and is given by

\begin{equation}
c^{j}_{\rm v}(t\vert R_{\rm V})=\frac{c_{\rm v}^{0}(j)\,\Omega_{j} z^{j}_{\rm v}\rho^{j}_{\rm d}(t)[2\gamma_{j}/R^{j}_{\rm V}(t\vert n)-\sigma_{j}]}{k_{\rm B}T[z^{j}_{\rm v}\rho^{j}_{\rm d}(t)+4\pi R^{j}_{\rm V}(t\vert n)\,c^{j}_{\rm V}(t\vert n)]}\ ,
\label{eqn-43}
\end{equation}
$\gamma_{j}$ is the surface tension of the void, $\sigma_{j}$ is the hydrostatic stress applied to the void and $c_{\rm v}^{0}(j)$ is the thermal-equilibrium vacancy concentration, given by

\begin{equation}
c_{\rm v}^{0}(j)=\frac{1}{\Omega_{j}}\exp\left[\frac{\Delta S^{\rm v}_{\rm f}(j)}{k_{\rm B}}\right]\exp\left[-\frac{E^{\rm v}_{\rm f}(j)}{k_{\rm B}T}\right]
=n_0(j)\,\exp\left[-\frac{E^{\rm v}_{\rm f}(j)}{k_{\rm B}T}\right]\ ,
\label{eqn-44}
\end{equation}
$\Delta S^{j}_{\rm f}$ is the change of entropy for the formation of point vacancy, and $E^{j}_{\rm f}$ is the point vacancy formation energy.
\medskip

By suing a continuous variable, the void concentration $c^{j}_{\rm V}(z,t\vert n)$ (with $n\geq 2$) introduced in Eqs.\,(\ref{eqn-40}) and (\ref{eqn-41}) can be obtained by solving the Fokker-Planck equation in the size space below (with $\xi=n$):

\begin{equation}
\frac{\partial c_{\rm V}^{j}(z,t\vert\xi)}{\partial t}={\cal G}^{j}_{0}(t\vert\xi)-\frac{\partial}{\partial\xi}
\left\{{\cal F}_{j}(z,t\vert\xi)\,c_{\rm V}^{j}(z,t\vert\xi)-\frac{\partial}{\partial\xi}\left[{\cal D}_{j}(z,t\vert\xi)\,c_{\rm V}^{j}(z,t\vert\xi)\right]\right\}\ ,
\label{eqn-45}
\end{equation}
where ${\cal F}_{j}(z,t\vert\xi)=z^{j}_{\rm v}\,D^{j}_{\rm v}\,c^{j}_{\rm v}(z,\,t)-z^{j}_{\rm i}\,D^{j}_{\rm i}\,c^{j}_{\rm i}(z,\,t)$ is the drift term, ${\cal D}_{j}(z,t\vert\xi)=[z^{j}_{\rm v}\,D^{j}_{\rm v}\,c^{j}_{\rm v}(z,\,t)+z^{j}_{\rm i}\,D^{j}_{\rm i}\,c^{j}_{\rm i}(z,\,t)]/2$ is the positive diffusion term, and ${\cal G}^{j}_{0}(t\vert\xi)$ is the cluster production rate per volume (with $\xi\geq 2$).
\medskip

Finally, the dislocation-loop density $\rho^{j}_{\rm d}(t\vert n)$ introduced in Eqs.\,(\ref{eqn-40}) and (\ref{eqn-41}) can be found from (with $n\geq 4$)

\[
\frac{\partial\rho^{j}_{\rm d}(t\vert n)}{\partial t}={\cal G}_{0}(z_j,t\vert n)+\left[\beta^{j}_{\rm v}(z_j,t\vert n+1)+\alpha^{j}_{\rm i}(z_j,t\vert n+1)\right]\rho^{j}_{\rm d}(t\vert n+1)
\]
\begin{equation}
+\beta^{j}_{\rm i}(z_j,t\vert n-1)\rho^{j}_{\rm d}(t\vert n-1)-\left[\beta^{j}_{\rm v}(z_j,t\vert n)+\beta_{\rm i}(z_j,t\vert n)+\alpha_{\rm i}(z_j,t,\vert n)\right]\rho^{j}_{\rm d}(t\vert n)\ ,
\label{eqn-46}
\end{equation}
where ${\cal G}_{0}(z_j,t\vert n)$ is the production rate per density for the interstitial dislocation loop of length $n$, the absorption ($\beta_{\rm i,v}$) and emission ($\alpha_{\rm i,v}$) rates in Eq.\,(\ref{eqn-46}) are defined by

\begin{equation}
\beta^{j}_{\rm i,v}(z_j,t\vert n)=2\pi\,\ell_j(n)\,z^{j}_{\rm c}(n)\,D^{j}_{\rm i,v}\,c^{j}_{\rm i,v}(z_j,\,t)\ ,
\label{eqn-47}
\end{equation}

\begin{equation}
\alpha^{j}_{\rm i,v}(z_j,t\vert n)=2\pi\,\ell_{j}(n)\,z^{j}_{\rm c}(n)\left[\frac{D^{j}_{\rm i,v}}{\Omega_j}\right]\exp\left[-\frac{E^{j}_{\rm b,\{i,v\}}(n)}{k_{\rm B}T}\right]\ ,
\label{eqn-48}
\end{equation}
$\ell_j(n)$ and $z^{j}_{\rm c}(n)$ are the radius and bias factor of an interstitial loop of size $n$, and $E^{j}_{\rm b}(n)$ is the binding energy for a cluster of $n$ interstitial atoms.
\medskip

The initial condition for the diffusion equations will be given by the corresponding calculated results from the atomic-scale model for individual layers. The point defect diffusion occurs mainly around interfaces between two adjacent layers or across the interfaces. The boundary conditions with continuous concentrations of point defects, as well as the jump in their derivatives determined by the dislocation sinks, will be applied at each interface. In addition, the constraints for the zero concentration of point defects as well as the zero derivative of the concentration with respect to $z$ at the two surfaces of the system will also be enforced in our numerical computations.
\medskip

\subsection{Point-Defect Electronic States}
\label{sec4.2}

To study the defect degradation effect on devices, we need to know not only the concentration and spatial distribution $\rho_{\rm d}({\bf r})$ of the irradiation-induced defects but also their electronic properties, such as energy level  $E_j$, wave function $\psi_j({\bf r})$, and local density of states ${\cal D}_{\rm d}({\bf r},\,E)$. Although the semi-classical MD calculation and the reaction-rate theory allow us to obtain the concentration and spatial distribution of defects, we still require density-functional theory\,\cite{book-estreicher,review} (DFT) for calculating defect configurations, energy levels, density of states and charge trapping by point defects in crystals.
\medskip

The main idea of DFT is to reformulate the energy of an atomic system as a functional of the ground state electron density function $\rho_0({\bf r})$ instead of individual electron wave functions. The proof of existence of such a functional relies on a one-to-one correspondence between the external potential $V_{\rm ext}(\{{\bf R}_\ell\},\,\{{\bf r}_m\})$ and $\rho_0({\bf r})$, where $\{{\bf R}_\ell\}$ and $\{{\bf r}_m\}$ label all the lattice atoms and electrons, respectively. The mapping of $V_{\rm ext}(\{{\bf R}_\ell\},\,\{{\bf r}_m\})$ onto $\rho_0({\bf r})$ is obvious. Any Hamiltonian $\hat{\cal H}$ with a given external potential $V_{\rm ext}(\{{\bf R}_\ell\},\,\{{\bf r}_m\})$ has a ground state solution with an $N$-electron wave function $\varphi_0(\{{\bf r}_m\})$, which can be uniquely associated with the electron density function $\rho_0({\bf r})$ using

\begin{equation}
\rho_0({\bf r})=N\int\cdots\int\left|\varphi_0({\bf r}_1,{\bf r}_2,\cdots,{\bf r}_N)\right|^2\,\delta({\bf r}-{\bf r}_1)\,d^3{\bf r}_1d^3{\bf r}_2\cdots d^3{\bf r}_N\ .
\end{equation}
Due to the resulting one-to-one correspondence between $V_{\rm ext}(\{{\bf R}_\ell\},\,\{{\bf r}_m\})$ and $\rho_0({\bf r})$, the energy $E_i$ of the atomic system can be expressed as a functional of the electron density $\rho_0({\bf r})$. The many-electron wave function of $\varphi({\bf r}_1,{\bf r}_2,\cdots,{\bf r}_N)$ depends on the `combination' of all spatial electron coordinates. Unfortunately, such an approach would by far exceed any computational capabilities. However, this problem can be overcome by using the Kohn-Sham (KS) ansatz,\,\cite{add11} in which the fully-interacting system is replaced by a non-interacting one. This approach corresponds to a mean-field approach, where the many-electron wave function is decomposed into a product of $N$ single-electron orbitals $\phi_i({\bf r})$ (i.e., Slater determinant). This simplification leads to a neglect of an energy contribution termed `correlations'. As a correction, the functional $E_{\rm xc}[\rho({\bf r})]$ must be introduced as an additional term in the Hamiltonian. Applying the variation principle to the modified Hamiltonian yields a single-particle-like Schr\"odinger equation, also referred to as Kohn-Sham equation in DFT. This equation includes an effective potential $V_{\rm eff}({\bf r})$, which is produced by the Coulomb forces of all other electrons and nuclei and incorporates the exchange and correlation interactions, i.e.,

\begin{equation}
\left[-\frac{\hbar^2}{2m_{\rm e}}\nabla^2+V_{\rm eff}({\bf r})\right]\phi_i^{\rm KS}({\bf r})=\varepsilon_i^{\rm KS}\phi_i^{\rm KS}({\bf r})\ ,
\label{dft2}
\end{equation}

\begin{equation}
V_{\rm eff}({\bf r})=V_{\rm ext}({\bf r})+V_{\rm ee}({\bf r})+V_{\rm xc}[\rho({\bf r})]\ ,
\end{equation}
where $V_{\rm ee}({\bf r})$ describes the electron-electron interaction (the classical Coulomb interaction) that is defined by

\begin{equation}		
V_{\rm ee}({\bf r})=\int d^3{\bf r}^\prime\,\frac{e^2\rho({\bf r}^\prime)}{4\pi\epsilon_0|{\bf r}-{\bf r}^\prime|}\ ,
\label{dft4}
\end{equation}
and $V_{\rm xc}[\rho({\bf r})]$ is the functional derivative of the exchange correlation energy with respect to the electron density function

\begin{equation}
V_{\rm xc}[\rho({\bf r})]=\frac{\delta E_{\rm xc}[\rho({\bf r})]}{\delta\rho({\bf r})}\ .
\label{dft3}
\end{equation}
The total energy of the atomic system can be arranged as

\begin{equation}
E[\rho]=T_k[\rho]+V_{\rm ext}[\rho]+V_{\rm ee}[\rho]+E_{\rm xc}[\rho]\ ,
\end{equation}
where $T_k[\rho]$ represents the kinetic energy of non-interacting electrons. The exchange-correlation functional in Eq.\,(\ref{dft3}) can be written as

\begin{equation}
E_{\rm xc}[\rho]=\left(T[\rho]-T_k[\rho]\right)+\left(V_{\rm el}[\rho]-V_{\rm ee}[\rho]\right)\ ,
\end{equation}
where $V_{\rm el}[\rho]$ is non-local electron-electron interaction beyond the classical one in Eq.\,(\ref{dft4}). $E_{\rm xc}[\rho]$ is simply the sum of the error made in using a non-interacting kinetic energy and the error made in treating the electron-electron interaction classically. The Kohn-Sham equations in Eq.\,(\ref{dft2}) have the same structure as the Hartree-Fock equations with the non-local exchange potential replaced by the local exchange-correlation potential $V_{\rm xc}[\rho({\bf r})]$. The computational cost of solving the Kohn-Sham equations scales formally as $N^3$ (due to the need to maintain the orthogonality of $N$ orbitals), but in current practice it drops to $N$ through the exploitation of the locality of the orbitals. Actually, the utility of the theory rests on the approximation made for $E_{\rm xc}[\rho]$.
\medskip

Therefore, the correct description of the exchange-correlation functional takes a crucial role in DFT. The local-density approximation has already achieved satisfactory results for systems with a slowly varying electron density function, such as metals.\,\cite{add12} However, it has a tendency termed over-binding, which overestimate binding energies and thus for instance predicts too strong hydrogen bonds with too short bonds lengths. The generalized gradient approximation is a systematic expansion, which gives good results in most cases, and corrects the issue of over-binding.\,\cite{add12} Recently, hybrid functionals\,\cite{add13} have emerged, which achieve an improved accuracy, especially for semiconductors with a bandgap.\,\cite{add14}
\medskip

Defect levels for charge capture or emission are calculated by means of the formation energies $E_f^q[X^{q^\prime}]$,\,\cite{add15} which are defined for a certain charge state $q$ and a certain atomic configuration $X^{q^\prime}$ of the defect as

\begin{equation}
E_f^q[X^{q^\prime}]=E_{\rm tot}[X^{q^\prime}]-E_{\rm tot}[{\rm bulk}]-\sum_j\,n_j\,\xi_j+q(\mu+\varepsilon_\nu+\Delta V)+E_{\rm corr}\ .
\end{equation}
Here, $E_{\rm tot}[{\rm bulk}]$ stands for the total energy of a super-cell containing pure bulk material while $E_{\rm tot}[X^{q^\prime}]$ represents the super-cell containing a defect. The third term corrects for the different numbers of atoms in both super-cells. The integer $n_j$ stands for the number of added ($n_j>0$) or removed ($n_j<0$) atoms which are required to create the defect from a perfect bulk structure, and $\xi_j$ denotes the corresponding energy in an atomic reservoir, which must be specified for each individual case. The fourth term accounts for the charge state $q$ of the defect, in which $\mu$ is defined as the electron chemical potential referenced with respect to the valence band edge $\varepsilon_\nu$ in a bulk-like region, and $\Delta V$ corrects the shift in the reference level between two differently charged super-cells and is obtained from the difference in the electrostatic potential far distant from the defect. Due to the periodic boundary conditions, charge neutrality must be maintained within a super-cell. Therefore, a homogeneous compensating background charge must be introduced in calculations of charged defects. This artificial Coulomb interaction is corrected by the last term $E_{\rm corr}$.

\subsection{Defect-Assisted Resonant Tunneling}
\label{sec4.3}

\begin{figure}[p]
\centering
\epsfig{file=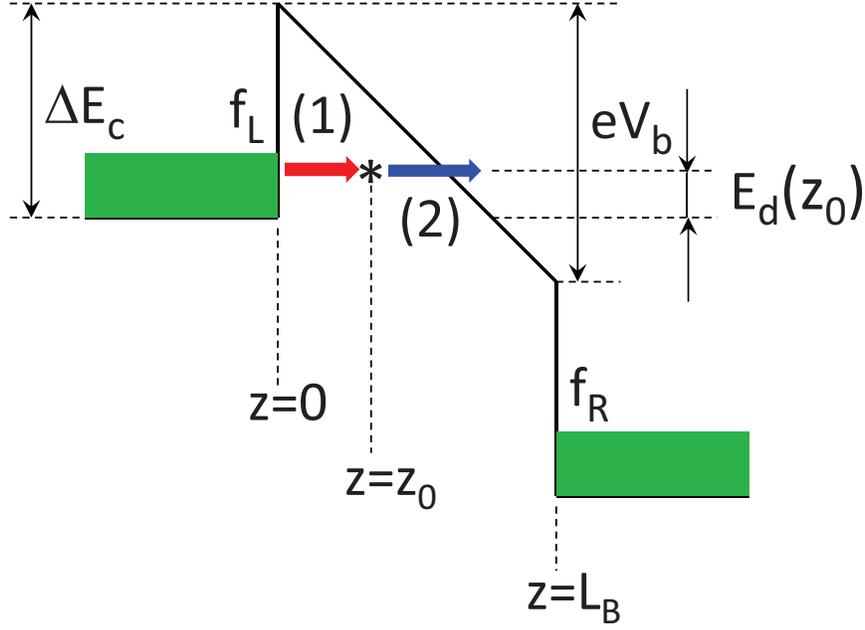,width=0.7\textwidth}
\caption{Schematic of defect-assisted resonant tunneling, where a point defect with energy $E=E_{\rm d}(z_0)$ at $z=z_0$ inside the barrier layer with conduction-band offset $\Delta E_{\rm c}$ and barrier thickness $L_{\rm B}$. The electron from the left electrode with Fermi distribution $f_{\rm L}$ is first captured (process-1 in red) by the point defect through tunneling, and then is emitted to a continuum state above the energy barrier (process-2 in blue)  through tunneling in the presence of a voltage drop $V_{\rm b}$ across the barrier layer.}
\label{f4}
\end{figure}

At low temperatures, the defect-assisted tunneling through thermal emission can be neglected.\,\cite{huang2} Therefore, the whole elastic tunneling  process can be divided into two subsequent ones, i.e., tunnel capture and tunnel emission, as shown in Fig.\,\ref{f4}. Although the in-plane momentum of electrons is not conserved during the tunneling process, the kinetic energy of electrons is conserved. For a neutral point defect, let us assume that it sits at an arbitrary position $z=z_0$ inside a barrier layer ($0\leq z_0\leq L_{\rm B}$) between the left ($L$) and right ($R$) electrodes with energy levels $0<E_{\rm d}(z_0)<\Delta E_{\rm c}$, where $\Delta E_{\rm c}$ is the conduction band offset for the middle barrier layer. A bias field ${\cal E}_{\rm b}$ is applied across the layer, leading to a voltage drop $V_{\rm b}={\cal E}_{\rm b}L_{\rm B}$.
\medskip

By assuming a large voltage drop, we need to consider only the forward current from left to right but not the backward current from right to left. In this picture, the left-going (capture) tunneling current density $J_{\rm L}(V_{\rm b},\,T)$ can be formally written as\,\cite{huang10}

\begin{equation}
J_{\rm L}(V_{\rm b},\,T)=2e\int_{0}^{L_{\rm B}} dz_0\, \rho_{\rm d}(z_0)\sum_{{\bf k}}\,\frac{2\pi}{\hbar}\left|\langle\Psi_{\bf k}\vert U_{\rm d}\vert\psi_{\rm d}\rangle\right|^2\,
\frac{\Gamma_{\rm d}/\pi}{[E_k-E_{\rm d}(z_0)]^2+\Gamma^2_{\rm d}}\,{\cal F}_{\rm L}(E_k)\ ,
\label{e54}
\end{equation}
where $\rho_{\rm d}(z_0)$ represents the distribution of point-defect concentration, $U_{\rm d}({\bf r})$ is the Coulomb potential associated with the point defect, $\psi_{\rm d}({\bf r})$ is its wave function, and $\Gamma_{\rm d}$ is the broadening in the density of states for point defects.
\medskip

In Eq.\,(\ref{e54}), the occupation factor ${\cal F}_{\rm L}(E_k)$ is defined as

\begin{equation}
{\cal F}_{\rm L}(E_k)=f_{\rm L}^{(0)}(E_k)\left\{1-g[E_{\rm d}(z_0)]\right\}{\cal Z}_{\rm e}-\left[1-f_{\rm L}^{(0)}(E_k)\right]g[E_{\rm d}(z_0)]\,{\cal Z}_{\rm f}\ ,
\label{e55}
\end{equation}
where $E_k=\hbar^2k^2/2m^\ast$ is the electron kinetic energy with effective mass $m^\ast$ in the left electrode, ${\cal Z}_{\rm e}$ and ${\cal Z}_{\rm f}$ represent the structural degeneracy factors of the point defect, when empty or filled, $g[E_{\rm d}(z_0)]$ is the defect occupancy function, and $f_{\rm L}^{(0)}(E_k)=\{1+\exp[(E_k-\mu_0)/k_{\rm B}T]\}^{-1}$ is the Fermi distribution function in the left electrode with chemical potential $\mu_0$. In addition, by employing the WKB approximation for the electron wave function $\Psi_{\bf k}({\bf r})$, the interaction matrix $\langle\Psi_{\bf k}\vert U_{\rm d}\vert\psi_{\rm d}\rangle$ is calculated as\,\cite{lannoo}

\begin{equation}
\langle\Psi_{\bf k}\vert U_{\rm d}\vert\psi_{\rm d}\rangle=\frac{A_k}{\sqrt{K(z_0)}}\,\exp\left[-\int_0^{z_0} dz^\prime\,K(z^\prime)\right]{\cal U}_1[K(z_0),\,{\bf k}_\|]\ ,
\label{e56}
\end{equation}
where $A_k$ is an unknown coefficient to be determined by the continuity of the wave function at the boundaries, ${\cal S}$ is the cross-sectional area,

\begin{equation}
K(z)=\frac{\sqrt{2m_2^\ast}}{\hbar}\,\left[\Delta E_{\rm c}-\frac{eV_{\rm b}z}{L_{\rm B}}\right]^{1/2}\ ,
\label{e57}
\end{equation}

\begin{equation}
{\cal U}_1[K(z_0),\,{\bf k}_\|]=\int d^3{\bf r}\,\psi_{\rm d}({\bf r})\,U_{\rm d}({\bf r})
\left[\frac{K(z_0)}{K(z)}\right]^{1/2}\,\frac{e^{-i{\bf k}_\|\cdot{\bf r}_\|}}{\sqrt{{\cal S}}}\,\exp\left[-\int^z_{z_0} dz^\prime\,K(z^\prime)\right]\ ,
\label{e58}
\end{equation}
${\bf r}=({\bf r}_\|,\,z)$, ${\bf k}=({\bf k}_\|,\,k_z)$.
\medskip

In a similar way, we can also calculate the right-going (escape) tunneling current density $J_{\rm R}(V_{\rm b},\,T)$. In the steady state, we have $J_{\rm L}(V_{\rm b},\,T)=-J_{\rm R}(V_{\rm b},\,T)\equiv J(V_{\rm b},\,T)$. This allows us to eliminate the unknown defect occupancy function $g[E_{\rm d}(z_0)]$ and eventually obtain\,\cite{lannoo}

\begin{equation}
J(V_{\rm b},\,T)=2e\,{\cal Z}_{\rm e}{\cal Z}_{\rm f}\left(f_{\rm L}^{(0)}-f_{\rm R}^{(0)}\right)\int_{0}^{L_{\rm B}} dz_0\, \rho_{\rm d}(z_0)\left[\frac{\Theta_{\rm R}}{{\cal P}_{\rm c}(z_0)}
+\frac{\Theta_{\rm L}}{{\cal P}_{\rm em}(z_0)}\right]^{-1}\ ,
\label{e59}
\end{equation}
where $f_{\rm R}^{(0)}(E_k)=\{1+\exp[(E_k-\mu_0+eV_{\rm b})/k_{\rm B}T]\}^{-1}$ and $\Theta_{\rm L,R}=f_{\rm L,R}^{(0)}{\cal Z}_{\rm e}+(1-f_{\rm L,R}^{(0)}){\cal Z}_{\rm f}$. In addition, the tunnel-capture rate (probability) ${\cal P}_{\rm c}(z_0)$ of an electron by a point defect in Eq.\,(\ref{e59}) is defined as

\begin{equation}
{\cal P}_{\rm c}(z_0)=\frac{2\pi}{\hbar}\sum_{{\bf k}}\,\left|\langle\Psi_{\bf k}\vert U_{\rm d}\vert\psi_{\rm d}\rangle\right|^2\,
\frac{\Gamma_{\rm d}/\pi}{[E_k-E_{\rm d}(z_0)]^2+\Gamma^2_{\rm d}}\ ,
\label{e60}
\end{equation}
and the tunnel-emission rate (probability) ${\cal P}_{\rm em}(z_0)$ of an electron captured by a strongly-localized point defect is given by

\begin{equation}
{\cal P}_{\rm em}(z_0)=\frac{e{\cal E}_{\rm b}}{4\sqrt{2m_2^\ast E_{\rm d}(z_0)}}\,\exp\left(-\frac{4\sqrt{2m_2^\ast E^3_{\rm d}(z_0)}}{3e\hbar{\cal E}_{\rm b}}\right)\ .
\label{e61}
\end{equation}
For photodetectors, the defect-assisted resonant tunneling greatly increases the dark current in the absence of incident light, which generates excess noise and reduces the detectivity of the photodetector.\,\cite{huang11}

\subsection{Reduced Carrier Mobility}
\label{sec4.4}

\begin{figure}[p]
\centering
\epsfig{file=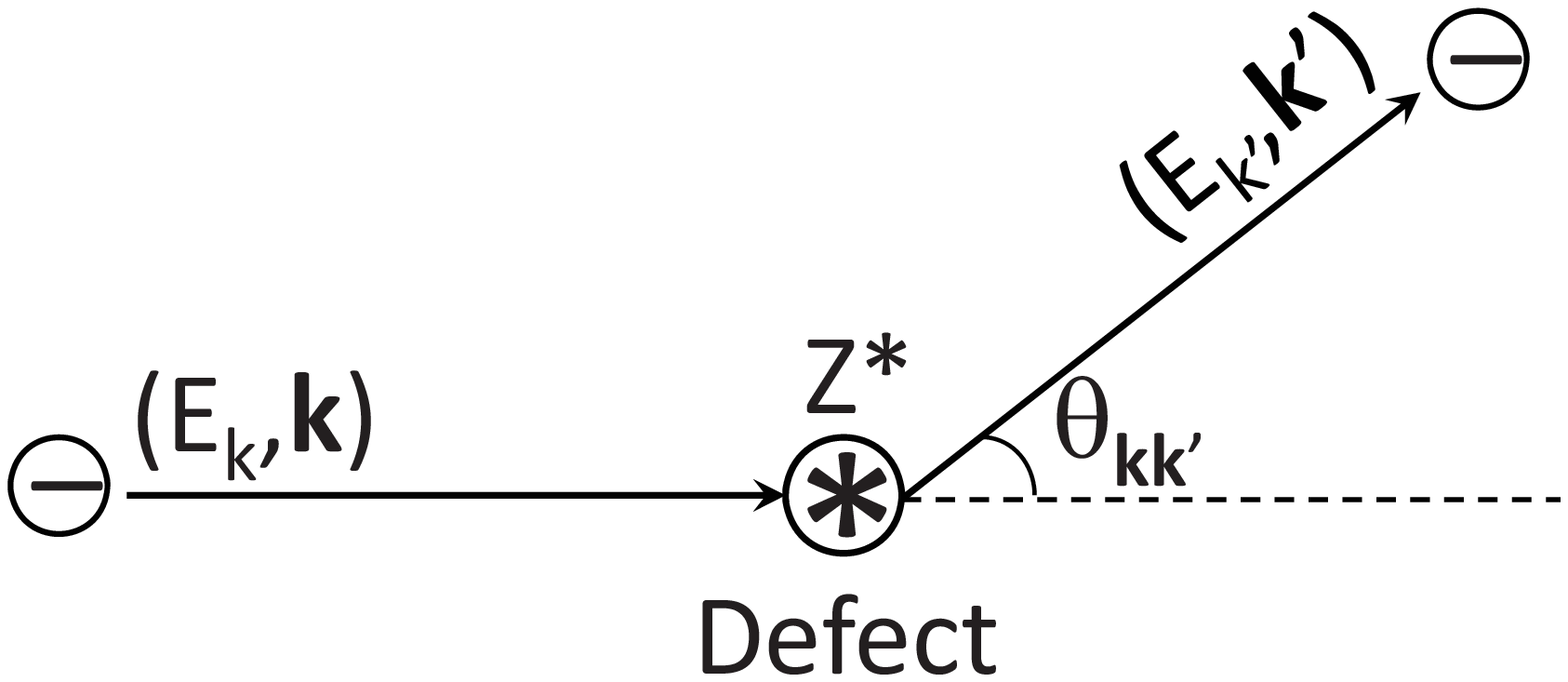,width=0.7\textwidth}
\caption{Schematic of scattering of an electron by a point defect, where the defect is charged and has an effective charge number $Z^\ast$. The incident electron with wave vector ${\bf k}$ and kinetic energy $E_k$ is scattered into a different direction with wave vector ${\bf k}^\prime$ and kinetic energy $E_{k^\prime}$. The scattering angle between ${\bf k}$ and ${\bf k}^\prime$ is denoted by $\theta_{{\bf k}{\bf k}^\prime}$ and the elastic scattering process requires $E_k=E_{k^\prime}$.}
\label{f5}
\end{figure}

When point defects are charged with a charge number $|Z^\ast|\geq 1$, they can scatter conduction electrons through their coulomb potential $\displaystyle{\sum_{i=1}^N \int d^3{\bf r}^\prime\,U_{\rm c}({\bf r}-{\bf r}^\prime)\,|\psi_{\rm d}({\bf r}^\prime-{\bf r}_i)|^2}$, as shown in Fig.\,\ref{f5}, where ${\bf r}_i$ for $i=1,\,2,\,\cdots,\,N$ represent the positions of $N$ point defects inside the quantum well and $\psi_{\rm d}({\bf r})\equiv\psi_{\rm d}({\bf r}_\|)\,\gamma_{\rm d}(z)$ is the wave function of the point defect in layered semiconductors. Let us consider electrons confined in one of the quantum wells with width $L_{\rm W}$ and barrier height $\Delta E_{\rm c}$. For simplicity, we assume that only the ground state of electrons is occupied at low temperatures with the wave function $\displaystyle{\Psi_{1{\bf k}_\|}({\bf r})=\frac{e^{i{\bf k}_\|\cdot{\bf r}_\|}}{\sqrt{{\cal S}}}\,\phi_1(z)}$ and subband energy $E_1(k_\|)=\varepsilon_1+\hbar^2k_\|^2/2m^\ast$ with quantum-well cross-sectional area ${\cal S}$, subband edge $\varepsilon_1$ and electron effective mass $m^\ast$.
\medskip

In this case, the interaction matrix $\langle\Psi_{1{\bf k}_\|}\vert U_{\rm c}\vert\Psi_{1{\bf k}^\prime_\|}\rangle$ is calculated as\,\cite{huang12}

\begin{equation}
\langle\Psi_{1{\bf k}_\|}\vert U_{\rm c}\vert\Psi_{1{\bf k}^\prime_\|}\rangle=\int dz\,\left|\phi_1(z)\right|^2\,\sum_{i=1}^N\,\sum_{{\bf q}_\|}\,U_{\rm c}(q_\|,\,z-z_i)\,
{\cal F}_{\rm d}(q_\|)\,\delta_{{\bf q}_\|,{\bf k}_\|-{\bf k}^\prime_\|}\,
e^{-i{\bf q}_\|\cdot{\bf r}_{i\|}}\ ,
\label{e62}
\end{equation}
where the two-dimensional Fourier transform of $U_{\rm c}({\bf r}-{\bf r}_i)$ is denoted as $U_{\rm c}(q_\|,\,z-z_i)$ and given by

\begin{equation}
U_{\rm c}(q_\|,\,z-z_i)=\left[\frac{\pm Z^\ast e^2}{2\epsilon_0\epsilon_{\rm r}(q_\|+q_s)}\right]\int dz^\prime\,e^{-q_\||z-z_i-z^\prime|}\,\left|\gamma_{\rm d}(z^\prime)\right|^2\ ,
\label{e63}
\end{equation}

\begin{equation}
{\cal F}_{\rm d}(q_\|)=\int d^2{\bf r}_\|\,e^{-i{\bf q}_\|\cdot{\bf r}_\|}\,\left|\psi_{\rm d}({\bf r}_\|)\right|^2\ ,
\end{equation}
$\epsilon_{\rm r}$ is the dielectric constant of the quantum-well host material. Moreover, $q_s$ in Eq.\,(\ref{e63}) is the inverse Thomas-Fermi screening length for quantum-well electrons, given by\,\cite{huang3}

\begin{equation}
q_s=\frac{e^2}{8\pi\epsilon_0\epsilon_{\rm r}k_{\rm B}T}\int_0^\infty dk_\|\,k_\|\,\cosh^{-2}\left[\frac{E_1(k_\|)-\mu_0}{2k_{\rm B}T}\right]\ ,
\end{equation}
where $T$ is the electron temperature and $\mu_0$ is the chemical potential of electrons in the quantum well.
\medskip

Since the positions of point defects are random, by introducing a continuous linear density distribution $\rho_{\rm 1d}(z_0)={\cal S}\rho_{\rm d}(z_0)$
for point defects, the interaction matrix from Eq.\,(\ref{e62}) becomes

\[
\left|\langle\Psi_{1{\bf k}_\|}\vert U_{\rm c}\vert\Psi_{1{\bf k}^\prime_\|}\rangle\right|^2=\left[\frac{Z^\ast e^2}{2\epsilon_0\epsilon_{\rm r}(|{\bf k}_\|-{\bf k}^\prime_\||+q_s)}\right]^2\,
\left|{\cal F}_{\rm d}(|{\bf k}_\|-{\bf k}^\prime_\||)\right|^2
\]
\begin{equation}
\times\int_{-L_{\rm W}/2}^{L_{\rm W}} dz_0\,\rho_{\rm 1d}(z_0)\left|\left(\int_{-\infty}^{\infty} dz\,\left|\phi_1(z)\right|^2\int dz^\prime\,
e^{-|{\bf k}_\|-{\bf k}^\prime_\|||z-z_0-z^\prime|}\,\left|\gamma_{\rm d}(z^\prime)\right|^2\right)\right|^2\ ,
\label{e64}
\end{equation}
where $L_{\rm W}$ is the width of quantum well. Once the scattering matrix elements in Eq.\,(\ref{e64}) are computed, by using Fermi's golden rule, the momentum-relaxation time $\tau_0$ can be obtained from

\begin{equation}
\frac{1}{\tau_0}=\frac{1}{2\pi\hbar}\int d^2{\bf k}_\|\int d^2{\bf k}^\prime_\|\,\left|\langle\Psi_{1{\bf k}_\|}\vert U_{\rm c}\vert\Psi_{1{\bf k}^\prime_\|}\rangle\right|^2\,
\delta[E_1(k_\|)-E_1(k^\prime_\|)]\left(1-\cos\theta_{{\bf k}_\|{\bf k}^\prime_\|}\right)\ ,
\label{e65}
\end{equation}
where $\theta_{{\bf k}_\|{\bf k}^\prime_\|}$ represents the angle between the two in-plane scattering wave vectors ${\bf k}_\|$ and ${\bf k}^\prime_\|$. By using the momentum-relaxation time $\tau_0$ in Eq.\,(\ref{e65}), the mobility $\mu_{\rm e}$ of electrons can be simply expressed as $\displaystyle{\mu_{\rm e}=\frac{e\tau_0}{m^\ast}}$. The reduced mobility of conduction carriers by radiation-induced point defects will directly affect the speed of high-mobility field-effect transistors in an integrated circuit.\,\cite{stern}

\subsection{Non-Radiative Recombination with Defects}
\label{sec4.5}

\begin{figure}[p]
\centering
\epsfig{file=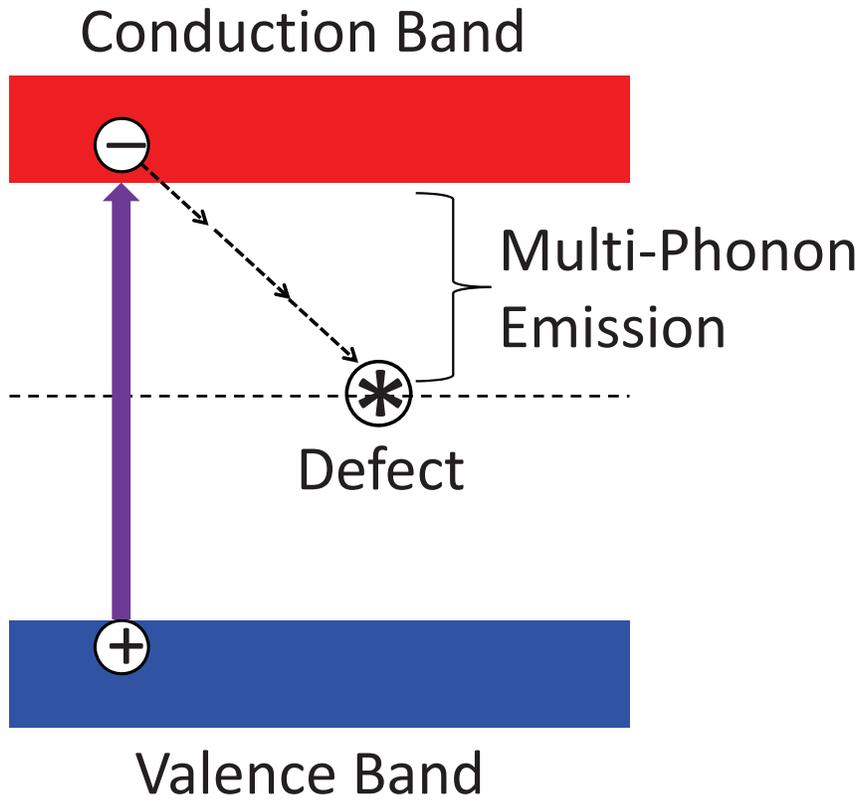,width=0.7\textwidth}
\caption{Schematic of non-radiative of photo-excited electrons from valence band to conduction band with point defects, where multiple phonons are emitted while the photo-excited electrons recombine with localized defect states within the bandgap.}
\label{f6}
\end{figure}

After the electrons are photo-excited from valence band to conduction band in a semiconductor, some of these photo-electrons will be quickly captured by point defects through an inelastic scattering process,\,\cite{huang10} as shown in Fig.\,\ref{f6}. By including the multi-phonon emission at room temperature\,\cite{huang6,ridley}, in this case the capture rate is calculated as\,\cite{multiphonon1,multiphonon2}

\[
{\cal W}_{\rm c}^{\rm e,h}({\bf k},\,z_0)=\frac{2\pi}{\hbar}\,\left|\langle\psi_{\rm d}\vert U_{\rm \{e,h\}p}\vert\Psi^{\rm e,h}_{\bf k}\rangle\right|^2\,\beta_{\rm HR}\left[1-\frac{\Delta E_{\rm e,h}(z_0)}{\hbar\Omega_0\beta_{\rm HR}}\right]^2\,
\]
\begin{equation}
\times\exp\left[-[2N_{\rm ph}(\Omega_0)+1]\beta_{\rm HR}+\frac{\Delta E_{\rm e,h}(z_0)}{2k_{\rm B}T}\right]\sum_{m=1}^{\infty}\,I_m(\xi)\,
\frac{\Gamma_{\rm d}/\pi}{[m\hbar\Omega_0-\Delta E_{\rm e,h}(z_0)]^2+\Gamma^2_{\rm d}}\ ,
\label{e68}
\end{equation}
where $U_{\rm \{e,h\}p}({\bf r})$ represent the potentials for the electron-phonon and hole-phonon coupling, $\psi_{\rm d}({\bf r})$ is the wave function of the point defect, $\Gamma_{\rm d}$ is the level broadening of the defect state, $\Psi^{\rm e,h}_{\bf k}({\bf r})$ are the wave functions of electrons (e) and holes (h) in a bulk, $\beta_{\rm HR}$ is the Huang-Ryhs factor, $\Delta E_{\rm e}(z_0)=E_{\rm G}+E^{\rm e}_k-E_{\rm d}(z_0)$, $\Delta E_{\rm h}(z_0)=E_{\rm d}(z_0)+E^{\rm h}_k$, $E^{\rm e,h}_k$ are the kinetic energies of electrons and holes, $E_{\rm G}$ is the bandgap energy of the semiconductor, $\hbar\Omega_0$ is the optical-phonon energy, $N_{\rm ph}(\Omega_0)=[\exp(\hbar\Omega_0/k_{\rm B}T)-1]^{-1}$ is the distribution function of thermal-equilibrium phonons, $T$ is the temperature, and $I_m(\xi)$ is the modified Bessel function of order $m$ with $\xi=2\beta_{\rm HR}\sqrt{N_{\rm ph}(\Omega_0)[N_{\rm ph}(\Omega_0)+1]}$.
\medskip

The electron-phonon coupling matrix element $|\langle\psi_{\rm d}\vert U_{\rm \{e,h\}p}\vert\Psi^{\rm e,h}_{\bf k}\rangle|^2$ in Eq.\,(\ref{e68}) can be evaluated by\,\cite{huang5}

\begin{equation}
\left|\langle\psi_{\rm d}\vert U_{\rm \{e,h\}p}\vert\Psi^{\rm e,h}_{\bf k}\rangle\right|^2=\frac{1}{(2\pi)^3}\,\int d^3{\bf q}\,\left|{\cal B}_{\rm d}({\bf q}-{\bf k})\right|^2
\left|U_{\rm \{e,h\}p}(q)\right|^2\ ,
\label{e69}
\end{equation}
where

\begin{equation}
{\cal B}_{\rm d}({\bf q}-{\bf k})=\int d^3{\bf r}\,\psi_{\rm d}({\bf r})\,e^{i({\bf q}-{\bf k})\cdot{\bf r}}\ ,
\end{equation}

\begin{equation}
\left|U_{\rm \{e,h\}p}(q)\right|^2=\frac{\hbar\Omega_0}{2}\left(\frac{1}{\epsilon_\infty}-\frac{1}{\epsilon_{\rm s}}\right)\,\frac{e^2}{\epsilon_0(q^2+Q_{\rm e,h}^2){\cal V}}\ ,
\label{e70}
\end{equation}
${\cal V}$ is the system volume,
$\epsilon_\infty$ and $\epsilon_{\rm s}$ are the high-frequency and static dielectric constants of the host semiconductor, and the inverse Thomas-Fermi screening length $Q_{\rm e,h}$ for bulk electrons and holes is given by

\begin{equation}
Q_{\rm e,h}^{2}=\frac{e^2}{\pi^2\epsilon_0\epsilon_{\rm r}k_{\rm B}T}\,
\int_0^\infty dk\,k^2\,f_0(E^{\rm e,h}_k-\mu_{\rm e,h})\left[1-f_0(E^{\rm e,h}_k-\mu_{\rm e,h})\right]\ .
\label{e71}
\end{equation}
Here, $f_0(E^{\rm e,h}_k-\mu_{\rm e,h})=\{1+\exp[(E^{\rm e,h}_k-\mu_{\rm e,h})/k_{\rm B}T]\}^{-1}$ is the Fermi distribution function for thermal-equilibrium conduction electrons and holes with chemical potentials $\mu_{\rm e,h}$.
\medskip

Finally, based on the given expression for $|\langle\psi_{\rm d}\vert U_{\rm \{e,h\}p}\vert\Psi^{\rm e,h}_{\bf k}\rangle|^2$ in Eqs.\,(\ref{e69})-(\ref{e70}),
the rate for the non-radiative recombination $\displaystyle{\frac{1}{\tau^{\rm e,h}_{\rm nr}}}$ can be explicitly calculated from

\begin{equation}
\frac{1}{\tau^{\rm e,h}_{\rm nr}}=\int dz_0\,\rho_{\rm 1d}(z_0)\left[\frac{1}{\tau^{\rm e,h}_{\rm d}(z_0)}\right]\ ,
\end{equation}

\begin{equation}
\left[\begin{array}{c}
1/\tau^{\rm e}_{\rm d}(z_0)\\
1/\tau^{\rm h}_{\rm d}(z_0)
\end{array}\right]
=\left[\begin{array}{c}
\left\{1-g[E_{\rm d}(z_0)]\right\}{\cal Z}_{\rm e}\,{\cal W}_{\rm e}(z_0)\\
g[E_{\rm d}(z_0)]{\cal Z}_{\rm f}\,{\cal W}_{\rm h}(z_0)
\end{array}\right]\ ,
\label{e72}
\end{equation}

\begin{equation}
{\cal W}_{\rm e,h}(z_0)=\int \frac{d^3{\bf k}}{(2\pi)^3}\,{\cal W}^{\rm e,h}_{\rm c}({\bf k},\,z_0)\,f_0(E^{\rm e,h}_k-\mu_{\rm e,h})\ ,
\end{equation}

\begin{equation}
{\cal W}^{\rm e,h}_{\rm c}({\bf k},\,z_0)={\cal R}_{\rm e,h}(z_0)\int \frac{d^3{\bf q}}{(2\pi)^3}
\left[\frac{e^2}{\epsilon_0(q^2+Q_{\rm e,h}^2)}\right]\left|{\cal B}_{\rm d}({\bf q}-{\bf k})\right|^2\ ,
\end{equation}

\[
{\cal R}_{\rm e,h}(z_0)=\frac{\pi\beta_{\rm HR}}{\hbar}\left(\frac{1}{\epsilon_\infty}-\frac{1}{\epsilon_{\rm s}}\right)e^{-[2N_{\rm ph}(\Omega_0)+1]\beta_{\rm HR}}\left[1-\frac{\Delta E_{\rm e,h}(z_0)}{\hbar\Omega_0\beta_{\rm HR}}\right]^2\exp\left[\frac{\Delta E_{\rm e,h}(z_0)}{2k_{\rm B}T}\right]
\]
\begin{equation}
\times\sum_{m=1}^{\infty}\,I_m(\xi)\,\frac{\Gamma_{\rm d}/\pi}{[m-\Delta E_{\rm e,h}(z_0)/\hbar\Omega_0]^2+\Gamma^2_{\rm d}}\ ,
\end{equation}
where $\rho_{\rm 1d}(z_0)={\cal S}\rho_{\rm d}(z_0)$ is the linear density distribution of point defects in a layer with the cross-sectional area ${\cal S}$, $g[E_{\rm d}(z_0)]$ in Eq.\,(\ref{e72}) is the defect occupancy function, and ${\cal Z}_{\rm e}$ and ${\cal Z}_{\rm f}$ represent the structural degeneracy factors of the point defect, when empty and filled, respectively. The steady-state condition for individual point defects requires $\displaystyle{\frac{1}{\tau_{\rm d}^{\rm e}(z_0)}=\frac{1}{\tau_{\rm d}^{\rm h}(z_0)}=\frac{1}{\tau_{\rm d}(z_0)}}$, which allows us to eliminate the unknown $g[E_{\rm d}(z_0)]$ introduced in Eq.\,(\ref{e72}), similar to what we have done in deriving Eq.\,(\ref{e59}). This leads to

\begin{equation}
\frac{1}{\tau_{\rm nr}}=\frac{1}{\tau^{\rm e}_{\rm nr}}=\frac{1}{\tau^{\rm h}_{\rm nr}}={\cal Z}_{\rm e}{\cal Z}_{\rm f}\int dz_0\,\rho_{\rm 1d}(z_0)\left[\frac{{\cal W}_{\rm e}(z_0){\cal W}_{\rm h}(z_0)}
{{\cal Z}_{\rm e}{\cal W}_{\rm e}(z_0)+{\cal Z}_{\rm f}{\cal W}_{\rm h}(z_0)}\right]\ .
\end{equation}
The change in the non-radiative time by point defects in the system will reduce the quantum efficiency of photo-excited electrons in both light-emitting diodes and photodetectors.\,\cite{huang13}

\subsection{Inelastic Light Scattering by Charged Defects}
\label{sec4.6}

Let us choose the $z$ direction perpendicular to the layered material. Light is incident on the layers in the $xy$-plane and scattered by charged
point defects within the layers. We consider an incident light with photon energy $\hbar\omega_{\rm i}$ and wave vector ${\bf k}_{\rm i}$ scattered inelastically by bound electrons within point defects at ${\bf r}_j=({\bf r}_{j\|},\,z_j)$ for $j=1,\,2,\,\cdots$. If the scattered-light photon energy and wave vector are denoted by $\hbar\omega_{\rm f}$ and ${\bf k}_{\rm f}$, respectively, the excitation energy and momentum transfer to charged point defects are given by $\hbar\omega=\hbar\omega_{\rm f}-\hbar\omega_{\rm i}$ and $\hbar{\bf q}=\hbar{\bf k}_{\rm f}-\hbar{\bf k}_{\rm i}$. We further assume that the ground and excited state (real) wave functions of defects are expressed as $\psi^{(0)}_{\rm d}({\bf r}-{\bf r}_j)=\psi^{(0)}_{\rm d}({\bf r}_{\|}-{\bf r}_{j\|})\,\gamma^{(0)}_{\rm d}(z-z_j)$ and $\psi^{(n)}_{\rm d}({\bf r}-{\bf r}_j)=\psi^{(n)}_{\rm d}({\bf r}_{\|}-{\bf r}_{j\|})\,\gamma^{(n)}_{\rm d}(z-z_j)$, where $n=1,\,2.\,\cdots$ represent different excited states of a charged point defect. The energy levels for the ground and excited states of charged defects are separately represented by $E^{(0)}_{\rm d}(z_j)$ and $E^{(n)}_{\rm d}(z_j)$.
\medskip

In a standard way, the differential scattering cross section $\displaystyle{\frac{d^2\sigma({\bf q},\,\omega)}{d\omega d\Omega_{\bf q}}}$ for inelastic light scattering can be shown to be\,\cite{huang7}

\[
\frac{d^2\sigma({\bf q},\,\omega)}{d\omega d\Omega_{\bf q}}=\left(\frac{e^2}{4\pi\epsilon_0m^\ast c^2}\right)^2\hbar\left|{\bf e}_{\rm i}\cdot{\bf e}_{\rm f}\right|^2\left(\frac{\omega_{\rm i}}{\omega_{\rm f}}\right)\frac{N_{\rm ph}(\omega)+1}{\pi}
\]
\begin{equation}
\times\int dz_0\,\rho_{\rm 1d}(z_0)\,{\rm Im}\left[\sum_{n,n^\prime}\,
{\cal A}_{nn^\prime}({\bf q}_\|,q_z\vert z_0)\,{\cal Q}_{nn^\prime}({\bf q},\omega\vert z_0)\right]\ ,
\label{e79}
\end{equation}
where ${\bf q}=({\bf q}_\|,\,q_z)$, ${\bf e}_{\rm i}$ and ${\bf e}_{\rm f}$ are the unit vectors for the polarizations of incident and scattered light, $N_{\rm ph}(\omega)=[\exp(\hbar\omega/k_{\rm B}T)-1]^{-1}$ is the photon distribution function, $T$ is the temperature, $\Omega_{\bf q}$ represents the solid angle in three-dimensional ${\bf q}$-space, and $\rho_{\rm 1d}(z_0)$ is the linear density of charged point defects.
\medskip

In addition, the form factor ${\cal A}_{nn^\prime}({\bf q}_\|,q_z\vert z_0)$ introduced in Eq.\,(\ref{e79}) is calculated as

\[
{\cal A}_{nn^\prime}({\bf q}_\|,q_z\vert z_0)=e^{-2iq_zz_0}\int d^2{\bf r}_\|\int d^2{\bf r}^\prime_\|\,\psi^{(n)}_{\rm d}({\bf r}_\|)\psi^{(0)}_{\rm d}({\bf r}_\|)\,\,e^{i{\bf q}_\|\cdot({\bf r}_\|-{\bf r}^\prime_\|)}\,\psi^{(n^\prime)}_{\rm d}({\bf r}^\prime_\|)\psi^{(0)}_{\rm d}({\bf r}^\prime_\|)\,
\]
\begin{equation}
\times\int dz\int dz^\prime\,\gamma^{({n})}_{\rm d}(z)\gamma^{({0})}_{\rm d}(z)\,e^{-iq_z(z+z^\prime)}\,
\gamma^{({n^\prime})}_{\rm d}(z^\prime)\gamma^{({0})}_{\rm d}(z^\prime)\ .
\end{equation}
The interacting density-density correlation function ${\cal Q}_{nn^\prime}({\bf q},\omega\vert z_0)$ employed in Eq.\,(\ref{e79}) is

\begin{equation}
{\cal Q}_{nn^\prime}({\bf q},\omega\vert z_0)=\sum_{m}\,\epsilon^{-1}_{nm}({\bf q},\,\omega)\,\Pi^{(0)}_{mn^\prime}(\omega\vert z_0)\ ,
\label{e81}
\end{equation}
where $\epsilon^{-1}_{nm}({\bf q},\,\omega)$ represents the matrix element of the inverse dielectric function of the host material containing defects.
In addition, $\Pi^{(0)}_{mn^\prime}(\omega\vert z_0)$ in Eq.\,(\ref{e81}) is the non-interacting density-density correlation function, given by

\begin{equation}
\Pi^{(0)}_{nn^\prime}(\omega\vert z_0)=\delta_{n^\prime,0}\,\frac{2n_{\rm d}^{(0)}(T\vert z_0)
\left[E_{\rm d}^{(n)}(z_0)-E_{\rm d}^{(0)}(z_0)\right]}
{\left[E_{\rm d}^{(n)}(z_0)-E_{\rm d}^{(0)}(z_0)\right]^2-\hbar^2\omega\left(\omega+i\Gamma_{\rm d}\right)}\ ,
\end{equation}
where we assume that only the ground state of charged point defects is occupied with the thermal occupation factor $n_{\rm d}^{(0)}(T\vert z_0)$.
\medskip

If there exist conduction electrons, in addition to bound electrons in charged point defects, with concentration $n_0$, effective mass $m^\ast$, and homogeneous broadening $\gamma_{\rm e}$ in the host material containing generated point defects, the matrix elements of the dielectric function are found to be\,\cite{huang8}

\begin{equation}
\epsilon_{nm}({\bf q},\,\omega)=\delta_{m,n}\left[1-\frac{\Omega^2_{\rm p}}{\omega(\omega+i\gamma_{\rm e})-3v_{\rm F}^2q^2/5}\right]\ ,
\label{e83}
\end{equation}
where $v_{\rm F}=(\hbar/m^\ast)\,(3\pi^2n_0)^{1/3}$ is the Fermi velocity of conduction electrons at zero temperature, $\Omega_{\rm p}=(n_0e^2/\epsilon_0\epsilon_{\rm b}m^\ast)^{1/2}$ is the plasma frequency, and $\epsilon_{\rm b}$ is the dielectric constant of the host material.
\medskip

Furthermore, if the host material is a doped polar semiconductor, its optical phonon modes can couple to conduction electrons. In this case, the matrix elements of the dielectric function in Eq.\,(\ref{e83}) are modified to\,\cite{huang9}

\begin{equation}
\epsilon_{nm}({\bf q},\,\omega)\rightarrow\delta_{m,n}\left\{1-\left[\frac{\omega(\omega+i\gamma_{\rm p})-\Omega^2_{\rm TO}}
{\omega(\omega+i\gamma_{\rm p})-\Omega^2_{\rm LO}}\right]\frac{\Omega^2_{\rm p}}{\omega(\omega+i\gamma_{\rm e})-3v_{\rm F}^2q^2/5}\right\}\ ,
\end{equation}
where the static dielectric constant $\epsilon_{\rm b}$ in the expression for $\Omega_{\rm p}$ should be replaced with the optical-frequency one $\epsilon_\infty$, $\Omega_{\rm LO}$ and $\Omega_{\rm TO}$ are the frequencies of the longitudinal and transverse optical phonon modes, and $\gamma_{\rm p}$
represents the phonon homogeneous broadening.
\medskip

The inelastic-light scattering technique can be used for identifying the charged point-defect species and their electronic properties,\,\cite{platzman} such as level separation between ground and excited states, broadening in the defect density of states, and optical polarization properties of point defects. If the incident coherent light is provided by a pulsed laser, the ultrafast dynamics of charged point defects can be directly measured and analyzed.\,\cite{tsen}

\section{Conclusions}
\label{sec5}

In conclusion, for the first time, we have proposed a multi-timescale microscopic model for fully characterizing the performance degradation of electronic and optoelectronic devices. In order to reach this goal, we have employed realistic interatomic potentials in a molecular-dynamics simulation for both the ultrafast displacement cascade stage and the intermediate defect stabilization and cluster formation stage. This simulation was then combined with a rate-diffusion theory for the slow defect reaction and migration stage. Additionally, with assistance from a density-functional theory for identifying defect species and their electronic properties, the calculated steady-state spatial distributions of defects and clusters were used to study and understand the physical mechanisms for the that degrade of electronic and optoelectronic devices, including defect-assisted resonant tunneling, reduced carrier mobility, non-radiative recombination with defects and inelastic light scattering by charged defects.
\medskip

In this paper, we have discussed several techniques for defect characterization. However, there are many other approaches for characterizing defect effects. These include electrical characterization techniques, such as deep-level transient spectroscopy and capacitance–voltage profiling, and optical characterization techniques, such as cathodoluminescence and reflectance modulation. Physical and chemical characterization techniques can also be applied, including electron energy loss spectroscopy,\,\cite{gumbs} secondary ion mass spectrometry\,\cite{weber} and chemical milling.
\medskip

The presented molecular dynamics model presented in this paper can be combined with a space-weather forecast model\,\cite{swf} which predicts a spatial-temporal flux of particle velocity distribution. With this combination of theories, according to the predicted irradiation condition for particular satellite orbits, electronic and optoelectronic devices can be specifically designed for operation in space with radiation-hardening techniques (such as self-healing and mitigation), which ensure components and systems are resistant to damage or malfunctions caused by particle and other types of radiation. This will effectively extend the lifetime of satellite onboard electronic and optoelectronic devices and greatly reduce the cost for long- or short-term space exploration. In addition, by improving the physical model for scintillation detectors, the accuracy of space-weather measurements will be enhanced.

\newpage

\end{document}